\documentclass[preprint]{aastex}
\usepackage{lscape,amsmath}
\usepackage{color,soul}
\pdfoutput=1

\slugcomment{Accepted for Publication in the Astrophysical Journal}

\shorttitle{Absorbers toward Lensed Quasars}
\shortauthors{Kulkarni et al.}

\begin{document}

\title{Probing Structure in Cold Gas at $z \lesssim 1$ with Gravitationally Lensed Quasar 
Sight Lines}

\author{Varsha P. Kulkarni}
\affil{University of South Carolina, Dept. of Physics and Astronomy, Columbia, SC 29208, USA}
\email{kulkarni@sc.edu}

\author{Frances H. Cashman}
\affil{University of South Carolina, Dept. of Physics and Astronomy, Columbia, SC 29208, USA}

\author{Sebastian Lopez}
\affil{Departamento de Astronom\'ia, Universidad de Chile, Casilla 36-D, Santiago, Chile}

\author{Sara L. Ellison}
\affil{University of Victoria, Dept. of Physics and Astronomy, Victoria, BC, V8W 2Y2, Canada}

\author{Debopam Som}
\affil{Department of Astronomy, The Ohio State University, Columbus, OH 43210, USA}

\author{Maria Jos\'e Maureira}
\affil{Max-Planck-Institut f\"ur extraterrestrische Physik, Giessenbachstrasse 1, 85748 Garching, Germany}

\begin{abstract}
Absorption spectroscopy of  gravitationally lensed quasars (GLQs) enables study of spatial variations in the interstellar  and/or circumgalactic medium of foreground galaxies. We report observations of 4 GLQs, each with two images separated by 0.8-3.0$\arcsec$, that show strong absorbers at redshifts 0.4 $<$$z_{abs}$$<$ 1.3 in their spectra, including some at the lens redshift with impact parameters 1.5-6.9 kpc. We measure H I Lyman lines along two sight lines each in five absorbers  (10 sight lines in total) using HST STIS, 
and metal lines using Magellan Echellette  or Sloan Digital Sky Survey. Our data have doubled the {lens galaxy} sample with measurements of H~I column densities ($N_{\rm  H I}$) and metal abundances along multiple sight lines. Our data, combined with the literature, show no strong correlation between absolute values of differences in $N_{\rm  H I}$, $N_{\rm  Fe II}$, or [Fe/H]  and the sight line separations at the absorber redshifts for separations of 0-8 kpc. 
The estimated abundance gradients show a tentative  anti-correlation with abundances at galaxy centers. Some lens galaxies show inverted gradients, possibly suggesting central dilution by mergers or infall of metal-poor gas. 
[Fe/H] measurements and masses estimated from GLQ astrometry suggest the lens galaxies lie below the total mass-metallicity relation for early-type galaxies { as well as measurements for quasar-galaxy pairs and gravitationally lensed galaxies} at comparable redshifts. { This difference may arise in part from the} dust depletion of Fe. { Higher resolution} measurements of  H and metals (especially undepleted elements) for more GLQ absorbers { and accurate lens redshifts} are needed to confirm these trends.\end{abstract}


\section{Introduction}
Imaging surveys form the cornerstone of our current understanding of galaxy evolution. 
These surveys indicate that the global star formation rate (SFR) in galaxies was 
considerably higher at $ z > 1$ than in the local universe, and that the bulk of the 
stellar matter in galaxies was assembled during $2 \la z \la 3$ (e.g., Madau \& Dickinson 2014). 
However, the majority of these surveys are designed to be flux-limited. 
A flux-independent probe  to study the evolution of the overall galaxy 
population is provided by absorption lines in quasar spectra superposed by intervening 
foreground galaxies,  sampled simply by 
gas cross-section, independent of their brightness. Quasar absorption systems, especially 
the damped Lyman-$\alpha$ absorbers (DLAs, defined as absorbers with H I column densities $\ge 2 \times 10^{20}$ 
cm$^{-2}$) and sub-DLAs (absorbers with $10^{19} \le N_{\rm H I} <  2 \times 10^{20}$) allow detailed 
investigations of the interstellar medium (ISM) in distant galaxies and the circumgalactic 
medium (CGM) around them, and trace evolution of metals and dust (e.g., Wolfe et al. 1986; P\'eroux et al. 2003; Pettini 2006, Kulkarni et al. 2005, 2007, 2010; Prochaska et al. 2007; Rafelski et al. 2012; Som et al. 2013, 2015; Quiret et al. 2016). 
This quasar absorption-line technique can, in principle, give a more representative 
view of distant galaxies than magnitude-limited imaging surveys. 

Traditional quasar absorption line studies probe a single sight line through a generally unidentified galaxy.  This represents two shortcomings: 1) it is difficult to connect the absorber's properties to the galaxy host, and 2) it is difficult to capture the galaxy's properties with a single sight line.  However, gravitationally lensed quasars (GLQs) can help to address both of these issues, since the lens galaxy may be seen in absorption (as well as other intervening absorbers), and one has multiple sight lines to characterize the galaxy and variations therein.  Observations of multiple sight lines have been successfully used in the Milky Way (MW) and nearby galaxies. 
 Indeed, the interstellar medium (ISM) within the MW and nearby 
galaxies shows structure over more than 8 orders of magnitude in spatial 
scale (e.g., Spangler 2001; McClure-Griffiths 2010). On small scales, the 
cold neutral gas (traced by Na I absorption lines) shows AU-scale variations caused by turbulence (e.g., Lauroesch et al. 2000; 
Andrews et al. 2001). At the same time, kpc-scale ISM structures 
such as supernova-driven bubbles are 
also evident. Such variations cannot be studied with absorption-line studies of single sight lines. 

In principle, sight lines to binary quasars also offer multiple probes of foreground galaxies. However, most binary quasars typically have separations in the 15-200 kpc range (e.g. Hennawi et al. 2006, Ellison et al. 2007; Green et al. 2011). In this study, we focus on GLQs rather than binary quasars because GLQs allow us to build a sample of sight lines  with close  ($<$10 kpc) transverse separations. Past observations of different kinds of absorbers toward GLQs 
have enabled constraints on absorber sizes and kinematics, using coincident and anti-coincident absorption 
features in 
the separate sight-lines (e.g., Smette et al. 1992; Rauch et al. 1999, 2001, 2002; Churchill 
et al. 2003; Ellison et al. 2004; Lopez et al. 2005, 2007; Chen et al. 2014; Zahedy et al. 2016, 2017; Rubin et al. {2018b}). Absorption studies of background galaxies and gravitationally lensed arcs are also becoming possible (e.g., Bordoloi et al. 2016, Rubin et al. { 2018a}, Lopez et al. 2018; P\'eroux et al. 2018). Closely separated sight lines to the background quasars or galaxies probe small-scale structure in the ISM of the lens galaxy. Furthermore, if other 
absorbers exist between the lens and the background source, structures on scales from very small 
(less than a parsec) to large (several kpc or tens of kpc, depending on the absorber 
location) can be probed, magnified by gravitational lensing. 

{The most sensitive probe of H I absorption is the Ly-$\alpha$ line, which lies in the UV for galaxies with $z < 1.6$. 
Given that most lensing galaxies are usually at $z$ $\la$ 1, their H I absorption studies require UV spectroscopy, as do the studies of any other $z < 1.6$ absorbers along the GLQ sight lines. Absorption studies of the lensing galaxies are especially important since the analyses of the GLQ images allow determinations of the galaxy mass, which can be combined with 
the gas properties to determine abundance gradients, constrain the mass-metallicity relation etc. Studies of non-lens 
absorbers toward GLQs are more common, but have been conducted primarily for $z > 1.6$ absorbers using optical spectra. 
The UV coverage and high spatial resolution of the Hubble Space Telescope (HST) allow measurement of H I absorption in both lensing galaxies and other absorbers at $z < 1.6$ along the sight lines to closely separated GLQ images.} With this in mind, we have obtained HST UV spectroscopy of 5 absorbers {at 0.4 $<$ $z_{abs}$ $<$ 1.3}   
toward 4 GLQs. This is a unique dataset since UV spectra exist for only a handful of GLQs (Michalitsianos et al. 1997; Lopez et al. 2005; Monier et al. 2009; Zahedy et al. 2017). 
Here we report results from these observations, along with the analysis of 
metal absorption lines based on Magellan Echellette (MagE) and Sloan Digital Sky Survey (SDSS) spectra. Section 2 describes the 
observations and data reduction. Section 3 describes line measurements and column density 
determinations. Section 4 presents a discussion of our results for element abundances 
along the individual sight lines, the implications for abundance gradients, a search for trends in absorption property differences with sight line separations, and constraints on the total mass vs. metallicity relation for lens galaxy absorbers. 
Section 5 summarizes our results and the outlook for future work. 
Throughout this paper, we adopt the cosmological parameters $\Omega_{m} = 0.3$, 
$\Omega_{\Lambda} = 0.7$, and $H_{0}$ = 70 km s$^{-1}$ Mpc$^{-1}$.

\section{Observations and Data Reduction}

{ \subsection{Targets} }

{ Our targets consist of 5 absorbers with $0.4 < z_{abs} < 1.3$ toward 4 GLQs. As mentioned in sec. 1, our main goal was to study absorption properties of the lens galaxies themselves. In some cases, the H I Lyman $\alpha$ or Ly-$\beta$ lines at the estimated lens redshifts could not be covered with STIS, but additional absorbers at other redshifts along the same sight lines could be covered. HST observations of these non-lens absorbers are also valuable given the paucity of GLQ absorbers at $z < 1.6$ with UV spectroscopy. At the time we designed our HST program, there were only 2 GLQs at similar angular separation that had been previously observed individually with HST in 2 separate programs, and only
1 of these had the absorber at the lens redshift.}

{  \subsubsection{Targets Observed}}

Table 1 lists basic properties of the GLQs in our sample, i.e., the number of images, quasar redshift, lens redshift, angular separation between the images, and the transverse separation between the GLQ sight lines at the lens redshift. Table 2 lists key properties of the targeted absorbers in these sight lines, i.e., the 
absorber redshift, rest-frame equivalent widths of the Mg II $\lambda$ 2796 absorption line in each sight line, the transverse separation between the sight lines at the absorber redshift, and the impact parameters of the lensed quasar images from the lens galaxies.  These GLQ sight lines with small angular separations allow us to probe small-scale structures.  
Fig. 1 shows HST or ground-based images of each field. 

The targets were selected from the CfA-Arizona Space Telescope LEns Survey (CASTLES; e.g., 
 Leh\'ar et al. 2000) and the SDSS Quasar Lens Search (SQLS) surveys,  with the following considerations:

\noindent $\bullet$ availability of some information about the lens redshift (photometric or spectroscopic); \\
\noindent $\bullet$ presence of {reasonably} strong Mg II absorption 
($W^{rest}_{2796}$$>$0.5 {\AA}) at coincident redshifts at the lens galaxy redshift or other redshifts 
$z_{abs}$$<$$z_{QSO}$ along multiple sight-lines based on already existing ground-based spectroscopy, indicating the need for UV spectroscopy ($z_{abs} < 1.6$); \\
\noindent $\bullet$ { at least 2 adequately UV-bright lensed images (as judged by the STIS exposure-time calculator)}.

 Given the Mg II equivalent widths, the sample absorbers  are good candidates to be DLAs or sub-DLAs, given the link between the 
strength of the Mg II absorption lines and the existence of a DLA/sub-DLA (e.g., Rao 
et al. 2006). This aspect of our sample selection strategy is important for metallicity determinations, since DLA/sub-DLAs enable accurate measurements of H I column densities and thus of metallicities. { [Thus, apart from requiring the presence of Mg II absorption, our survey is blind to the other galaxy properties. It is possible that stronger Mg II absorption tends to favor more luminous galaxies, in which case less luminous galaxies may be less likely to show Mg II absorption in sight lines separated by several kpc. On the other hand, some galaxies, despite being associated with strong  Mg II absorption, are found to have low luminosity (e.g., P\'eroux et al. 2011a and references therein). It is still debated whether or not MgII selection could introduce a metallicity bias (e.g., Kulkarni et al. 2007, 2010; Dessauges-Zavadsky et al. 2009; Som et al. 2015; Berg et al. 2016). In any case, in this paper, we are using the dual lines of sight to look at differences within a galaxy.  Our interest is in merely the relative values of the gas column densities and metallicities in the two sight lines, not the actual absolute values of these quantities. Therefore, we emphasize that any potential effect of Mg II selection is not relevant to the main goal of our program.]}  
 
{  \subsubsection{Past Observations}
 
 We now briefly review the properties of the GLQs studied in this work based on past observations. One of the most important properties is the redshift of the lens galaxy. Ideally, the lens galaxy redshifts for all the targets would come from the same method.  Unfortunately such a uniform procedure is not possible for our sample, since the different GLQs have been observed in different studies with different quality of observations and thus different degrees of reliability. We have therefore used our judgment and the quality of the data available for
each object to decide on a case-by-case basis which redshift is likely to be more accurate.
 } 
 
 Q1017-2046 was reported to be a gravitationally lensed quasar at $z=2.545$ with an image separation of 0.85$\arcsec$  
 (Surdej et al. 1997). The lens galaxy was not detected in HST WFPC2 V and I band images of Remy et al. (1998), but was 
 detected in HST NICMOS H-band images of Lehar et al. (2000). The redshift of the lens galaxy was estimated to be 0.78 from the available photometry (Kochanek et al. 2000), but was suggested to be $1.088 \pm 0.001$ based on the presence of strong Mg II absorption 
 features in both sight lines  (Ofek et al. 2006). Our data confirm the presence of { absorption lines of H I and several metal ions in both sight lines in the range  $z=1.086-1.089$}, and we adopt the redshift of the dominant metal component (1.086) as the lens redshift. 
 
 Q1054+2733 was discovered as a new gravitationally lensed quasar with an image separation of 1.27$\arcsec$ in the SDSS by Kayo et al. (2010), who estimated the redshift of the lensing galaxy to be about 0.23, based on the galaxy photometry and the Faber-Jackson relation. A strong Mg II absorber at $z=0.68$ was detected in the sight lines of Q1054+2733A, B (Kayo et al. 2010, Rogerson \& Hall 2012). Our HST STIS spectra do not cover the H I Ly-series lines for the $z=0.23$ 
 lens, but are targeted for the $z=0.68$ absorber. 
 
 Q1349+1227 was 
  shown to be a gravitationally lensed quasar with an image separation of 3.00$\arcsec$ by Kayo et al. (2010), who reported the detection of a faint lens galaxy close to the faint quasar image, and estimated the redshift of the lensing galaxy to be about 0.63, based on the galaxy photometry and the Faber-Jackson relation.  An Mg II and Fe II absorber was reported at $z=1.24$ (Kayo et al. 2010; Quider et al. 2011). Our HST STIS spectra do cover the H I Ly-$\alpha$ feature for the lens, but are very noisy due to the low continuum flux level caused by the presence of the Lyman break for the absorber at $z=1.24$. 
 
 Q1355-2257 (CTQ 327) was 
 shown to be gravitationally lensed with an image separation of 1.22$\arcsec$  based on 
 HST STIS images by Morgan et al. (2003). The lens galaxy was estimated to be an early-type galaxy with a redshift in the range of 0.4 to 0.6, on the basis of the galaxy photometry in I and K bands, and the Faber-Jackson relation by Morgan et al. (2003). 
 Mg II absorption is detected at $z=0.48$, and a tentative detection of a 4000 {\AA} break at $z=0.48$ has also been reported (Ofek et al. 2006). { Eigenbrod et al. (2006) attempted to obtain a spectrum of the lensing galaxy from spatial deconvolution of the quasar spectra. However, they could not obtain the lens redshift securely owing to low S/N resulting from the close proximity ($< 0.3 \arcsec$) of the galaxy to the quasar image B, and provided only a tentative estimate of 0.70.} Eigenbrod et al. (2007) detect Mg II absorption at $z=0.70$ in sight line B, but not A, and suggest $z=0.70$ as the lens redshift. Chen et al. (2013) report X-ray observations of Q1355-2257, but detect no Fe K $\alpha$ emission in image A or B. Our data confirm the presence of H I and metal absorption in both sight lines at $z=0.48$. We also confirm the detection of H~I absorption at $z=0.70$ in both sight lines, and the detection of metal absorption in sight line B (but not sight line A) at $z=0.70$.  It is possible that the metal absorber at $z=0.70$ is associated with a galaxy other than the lens galaxy whose Mg II cross-section is just not large enough to cover both sight lines. The presence of H I and metal absorption at $z=0.48$ in both sight lines also makes it a { potential} candidate for the lens galaxy. In the analysis below, we determine the abundance gradients for both scenarios, considering the lens galaxy redshift to be (a) $z=0.48$ and (b) $z=0.70$. 
 
 \subsection{HST Spectra} 
HST observations were carried out with the Space Telescope Imaging Spectrograph (STIS) with the goal of measuring the H I Ly-$\alpha$ (and, in most cases, also Ly-$\beta$) lines in the absorbers of interest. 
STIS was chosen over COS in order to efficiently obtain spatially resolved long-slit 
spectra of the individual GLQ images. The two images for each GLQ were aligned along the cross-dispersion direction of the 52"x0.2" STIS slit. The optimum slit orientations were calculated using the known RA and DEC separations of the lensed images, and ORIENT ranges of 2 degrees centered on the optimum value needed to line up both images in the slit were chosen. The G230L grating { (which has a spectral resolution of $R \sim 500$)} was used with the NUV-MAMA to cover the Ly-$\alpha$ and Ly-$\beta$ absorption lines. Fig. 1 indicates the orientation of the HST STIS slit for our observations. Table 3 summarizes the observations.                                                                          
                                                                                    
Each GLQ pair was acquired with onboard target acquisition. No other brighter sources exist within the 5$\arcsec$$\times$5$\arcsec$ aperture from the brighter quasar image in each pair of sight lines. For each orbit, two spectroscopic exposures were obtained, corresponding to two positions of a STIS-ALONG-SLIT dither pattern. A total of 4 dither patterns were employed,  each  with two positions separated by 1.5$\arcsec$, 2.0$\arcsec$, 2.5$\arcsec$, or 3.0$\arcsec$, so as to prevent the spectra of the lensed images in different orbits from falling on the same part of the detector.   
     
The HST STIS spectra were first reduced using the IRAF/STSDAS task CALSTIS. The CALSTIS pipeline reduction consists of overscan subtraction, bias subtraction, cosmic ray rejection, linearity correction, dark subtraction, flat-fielding, processing of the 
contemporaneously obtained wavelength calibration data, wavelength calibration, flux calibration, and 2-D rectification. The reduced 2-dimensional data were examined to determine the locations of the two quasar images, and further processed using the X1D task to extract 1-dimensional spectra. The X1D task also performs background subtraction, charge transfer efficiency correction, conversion to heliocentric wavelengths, and absolute flux calibration. The extracted 1-dimensional spectra for the individual sub-exposure were co-added to calculate the combined spectra, including the flux uncertainties. Continuum fitting was performed with the IRAF task CONTINUUM using low-order cubic spline 
polynomial functions, and the continuum-normalized spectra were examined to measure the H I absorption lines. 

While the low-resolution STIS spectra are adequate for measuring the H I Lyman absorption lines in our sample, they are not suitable for measurements of the much narrower metal absorption lines. The metal lines were measured with either Magellan spectra or SDSS spectra. { While the spectral resolution of our MagE data is moderate ($\sim$51 km s$^{-1}$), we note that the high S/N of these data makes them adequate for deriving reliable metal column densities. Indeed, a number of past studies of element abundances for quasar absorbers have been based on spectra with resolution comparable to or lower than our MagE spectra (e.g., 44 km s$^{-1}$ for Keck ESI, 75 km s$^{-1}$ for MMT Blue Channel). In cases where both high S/N medium-resolution spectra and high-resolution spectra are available for the same object, the column densities derived from the two sets of spectra are found to agree well (e.g., comparison of Keck ESI and HIRES results in Prochaska et al. 2007, and comparison of MMT blue channel and VLT UVES results in P\'eroux et al. 2006).}

\subsection{Magellan Spectra} 
The quasars Q1017-2046 and Q1355-2257 were observed using the Magellan Echellette (MagE) spectrograph 
on the Clay 6.5 meter Magellan telescope at the Las Campanas Observatory in  March 2010, with the goal of measuring metal absorption lines (e.g., lines of Mg I, Mg II, Fe II). The excellent seeing { (0.5-0.7")} allowed 
observations with a 0.7" slit{, giving a spectral resolution of $R \sim 5900$}. Multiple exposures lasting 2700-3600 seconds each were taken separately for each of the quasar 
images. { In each case, the slit was aligned with the parallactic angle in order to minimize chromatic slit losses, since the
MagE spectrograph does not have an Atmospheric Dispersion Corrector. The spectrum of the fainter images thus contained some light 
from the brighter image. This was corrected for in the extraction as follows:  Two Gaussians + a sky floor were fit simultaneously, with all parameters free. In a second fit, we tied the Gaussian positions and common width by a model that we obtained from fitting those measurements as a function of wavelength with a smooth polynomial. In the second double Gaussian fit only the amplitudes and the sky were free to vary. This gave a  much more robust fit. The (barely) resolved fluxes were the areas of the fitted Gaussians. More details of 
this procedure can be found in Lopez et al. (2005).}

The MagE data were reduced using { custom} FORTRAN routines which interact with MIDAS commands (Ballester et al. 1992). The MIDAS task ``IDENTIFY/ECHELLE" was used 
to calibrate the wavelengths using the 2D Th-Ar lamp spectrum. The wavelengths were converted to vacuum wavelengths and 
heliocentric corrections were calculated using ``COMPUTE/BARYCORR". The calibrated Echelle orders were continuum-fitted individually using cubic 
spline polynomials. The normalized orders were then merged into a single spectrum.

{ \subsection{Keck Spectra}

For one of our GLQs (Q1017-2046), archival Keck HIRES spectra are available. Although these spectra are not spatially resolved, their higher spectral resolution makes them useful to obtain a sanity check on the metal column densities estimated from the MagE data. Extracted HIRES spectra for each order containing the lines of interest were downloaded from the Keck Observatory Archive and dispersion-corrected to a dispersion of 0.075 {\AA} to align the exact starting wavelengths for the individual exposures of a given order. Continuum-fitting was performed with the PyRAF task ``CONTINUUM" using cubic spline polynomials of orders 2 to 5. The individual exposures for each order of interest were co-added using the PyRAF task ``SCOMBINE'' (three exposures for Mg II, four exposures for Fe II).} 

 \subsection{SDSS Spectra} 
For Q1054+2733 and Q1349+1227, we do not have MagE { or HIRES} data, and use spectra from the Sloan Digital Sky Survey (York et al. 2000), that were originally part of Data Release 7 (Abazajian et al. 2009) to examine the metal lines. These spectra are not resolved (as the lensed image separations for these GLQs are smaller than or equal to the SDSS fiber diameter), and thus do not allow metal line measurements in the individual sight lines. Nevertheless, the SDSS spectra  still give a useful indication of the integrated metal line strengths. The spectra 
were split into multiple sections of $\sim30-500$ {\AA} to facilitate 
continuum fitting. 
The pieces were continuum-fitted with the PyRAF task ``CONTINUUM". Cubic spline polynomials of order $\thicksim2-10$ 
were typically the best fit for both quasar spectra, as they resulted in the lowest root mean square (rms) residuals. The normalized 
split sections for each quasar were then combined into a single spectrum using the PyRAF task ``SCOMBINE" for subsequent 
analysis.

\subsection{Absorption Line Measurements and Column Density Determinations} 

The H I and metal lines were identified and measured in each 
sight line. Column densities were measured using the Voigt profile fitting software VPFIT (https://www.ast.cam.ac.uk/$\sim$rfc/vpfit.html). 

{ \subsubsection{H I Lines}}
Table 4 lists the equivalent widths of the Ly-$\alpha$ lines, and of the Ly-$\beta$ lines where available. We note, however, that in some cases (e.g., Q1017A, B $z_{abs} = 1.09$ and Q1355A $z_{abs} = 0.70$), it is difficult to measure the Ly-alpha and/or Ly-$\beta$ equivalent widths accurately due to blends or the presence of nearby features (arising in metal lines, Ly-alpha forest lines, and/or Milky Way interstellar lines), and the presence of noise. { It is thus not possible to reliably estimate 
 the H I column densities from just the equivalent widths assuming a theoretical profile. We therefore estimate $N_{\rm H I}$ by fitting  Voigt profiles, convolved with the instrumental profile, to the data.} A procedure similar to that of Rao \& Turnshek (2000) and Rao et al. 
(2006) was followed while estimating the H I column densities from the HST spectra. Voigt profile fitting was performed for the absorption lines of Ly-$\alpha$ (and Ly-$\beta$ where available and not severely affected by noise or blending with other Ly-$\alpha$ forest lines), including the convolution with the STIS G230L line  
spread function provided by the Space Telescope Science Institute. (Higher order Lyman lines were not available.) Given the low 
resolution of the G230L data, the relatively low S/N in some spectra, and the presence of 
other nearby lines, it is not meaningful to vary all three 
parameters (i.e., H~I column density, Doppler $b$ value, and the redshift)  while performing a Voigt profile fit {  to the Ly-$\alpha$ and/or Ly-$\beta$ lines in the absorbers of interest}. { Instead, the H I column density and redshift were allowed to vary and,} the Doppler $b$ parameter 
was {  fixed during the profile fits.} { (We note that the difference between the H I and metal-line redshifts is less than the velocity resolution of the G230L, and repeating the H I Voigt profile fits using the redshifts derived from the metal lines made $< 1 \sigma$ difference to the H I column densities determined.)} {  The fits were repeated for a variety of fixed $b$ values ranging from 15 km s$^{-1}$ to the maximum value deemed possible given the Doppler $b$ parameters of the metal lines, if available, for the individual sight lines. The range of $b$ values explored varied between 15-30 and 15-87 km s$^{-1}$.} 
  {  If the $N_{\rm H I}$ values for the maximum and minimum $b$ values differed, the average of the two values was adopted as the $N_{\rm H I}$ value, and the sampling error in this average value was added in quadrature with the measurement uncertainties to obtain the total $N_{\rm H I}$ uncertainty. Table 5 lists the adopted $N_{\rm H I}$  values and uncertainties.  For the $z_{abs}=0.70$ system toward Q1355-2257A (for which no metal lines are detected),  we adopt the $N_{\rm H I}$  value (log $N_{\rm H I}$=$18.20 \pm 0.23$) for $b=15$ km s$^{-1}$
  and note that $N_{\rm H I}$ values within $\lesssim$ 1.5 $\sigma$ of this adopted  value are obtained for a reasonable range of $b$ values (15-30 km s$^{-1}$). Figs. 2-6 show the H I profile fits for the best-fit and $\pm 1 \sigma$ $N_{\rm H I}$ values. In cases where the  $N_{\rm H I}$ values obtained for the minimum and maximum $b$ values differ by $\ge 0.02$ dex, the fits for both sets of values are shown.}

For the $z=0.68$ absorber toward Q1054+2733B, 
{ the noisy STIS data barely show a weak Ly-$\alpha$ line} that is $\sim$ 1.2 resolution element away (730 km s$^{-1}$ away) from the expected Ly-$\alpha$ position for the redshift based on the Mg II lines seen in the unresolved SDSS spectrum. It is possible that this feature may not be the Ly-$\alpha$ associated with the $z=0.68$ Mg II absorber. Given the uncertainty in the detection of this feature in the noisy spectrum, we conservatively adopt an upper limit on the H I column density for this sight line estimated from the 3 $\sigma$ sensitivity for the Ly-$\alpha$ line (including contributions to the equivalent width uncertainty from both photon noise and 
 continuum determination uncertainty, and assuming the damping part of the curve of growth to get a conservative upper limit). 

 To summarize, { we adopt the H I column densities listed in Table 5. Follow-up higher S/N and higher resolution spectra would help to obtain more definitive H I column densities.  The highest H~I column densities are found along the sight lines to Q1349+1227A, B.}

{ \subsubsection{Metal Lines}}
For the metal lines, where multiple lines of the 
same ion were covered 
by the data, they were used together to constrain the ionic column densities. The Doppler $b$ parameters and redshifts were tied together for 
all low ions (e.g., Si II, Fe II etc.). 
The total column densities were calculated by adding all the velocity components 
together. Furthermore, to ensure the reliability of the results, the column densities of the unsaturated and unblended lines 
were also measured by the apparent optical depth method (AOD; Savage \& Sembach 1991) and compared to the measurements from Voigt 
profile fitting. As another check of the VPFIT results, we also performed single-component curve-of-growth analyses for all available Fe II lines [using both the Python package linetools (http://linetools.readthedocs.io/en/latest/index.html), and a comparison of the observed data with theoretical curves-of-growth for a range of $b$ parameters].  
The Fe II column densities and effective $b$ parameters estimated from the curve-of-growth analyses agree closely with those estimated from VPFIT.

The uncertainties in the effective Doppler $b$ parameters listed in the tables are those estimated from VPFIT, and 
are typically $\lesssim 20\%$. Details about the estimation and reliability of parameter uncertainties in VPFIT can be found in the 
VPFIT manual (Carswell \& Webb 2014). We independently checked these uncertainties using single-component curve-of-growth analyses for the available Fe II lines, and obtained uncertainties close to those returned by VPFIT, typically  within 1-2 km s$^{-1}$. Additionally, we performed single-component curve-of-growth analyses for cleanly isolated individual components of the Fe~II lines where possible, and once again obtained uncertainties close to those estimated from VPFIT. 
It is still possible that in some cases the uncertainties in the $b$ parameters may be underestimates. In any case, the metal column densities estimated from VPFIT show generally good agreement (within $\lesssim 1 \sigma$) with the AOD values and the curve-of-growth estimates. Oscillator strengths from Cashman et al. (2017) were adopted for use in all metal-line column density  
measurements.

\section{Results}

\subsection{Primary Absorption Systems}

Metal lines for a number of ions (e.g., Mg I, Mg II, Fe II, Si II, Al II, and/or Al III) were measured in the Magellan and SDSS spectra. Figures  \ref{fig_1017a_pro} - \ref{fig_1355b_pro2} show the profile fits to the 
metal lines in the absorbing systems. For Q1017-2046 and Q1355-2257, the resolved MagE spectra of images A and B allow profile fitting of the metal lines in each sight line. For Q1054+2733 and Q1349+1227, the unresolved SDSS spectra allow metal line fits for  the combined A+B image. Tables 6 and 7 list the metal column density results from Voigt profile fitting for the $z=1.09$ absorber toward Q1017-2046A and Q1017-2046B, respectively. Tables 8 and 9 list, for both of these sight lines, the total { metal} column densities (summed over all the detected velocity components) from the profile fits and the AOD method where available{, based on the MagE data. Also listed are the corresponding element abundances derived from the metal and H I column densities. Tables 10 and 11 list the corresponding values obtained from the spatially unresolved Keck HIRES spectra of Q1017-2046A+B. While these values cannot be directly compared to the  values for the individual sight lines from the MagE data, it is reassuring that the sum of the MagE values for Fe II column densities for sight lines A and B are consistent with the HIRES values for Q1017-2046A+B. (There is a bigger difference for Mg II, which may be partly due to a higher degree of saturation. In any case, none of our discussion on abundance gradients in sec. 4 is affected by these differences, since it is based on Fe II, for which both HIRES and MagE data give about the same total column density.)} 

{ We also note that, on degrading the Q1017-2046A+B HIRES data to the MagE resolution, we find column densities derived for the latter data to be within 0.02-0.05 dex of the values obtained from the original HIRES data. Thus, overall, we regard our metal column density estimates from the MagE data as reasonably robust. In the two cases where we do not have MagE or HIRES spectra and are forced to use SDSS spectra (Q1054+2733A,B and Q1349+1227A,B), the metal lines are not resolved; in these cases, the AOD column densities are likely to be more accurate than the VPFIT estimates.}

Tables 12-14 and Tables 15-17 list the metal column density and element abundance results for the unresolved images of Q1054+2733 and Q1349+1227, respectively. Tables 18 and 19 list the metal column density results from Voigt profile fitting for the $z=0.48$ absorber toward Q1355-2257A and Q1355-2257B, respectively. Tables 20 and 21 list the total column densities, and the element abundances for both of these sight lines.  { Tables 24 and 25 list the  metal column densities and element abundances for the $z=0.70$ absorber toward Q1355-2257B.} No metal lines are detected at $z=0.70$ toward Q1355-2257A; { Tables 22 and 23 list the corresponding 3 $\sigma$ upper limits on the metal column densities and on the abundances.} 

In all cases, the total abundances for the available elements (Tables { 9, 14, 17, 21, 23}) were derived from the total column densities for the respective elements, adopting solar 
photospheric abundances from Asplund et al. (2009) as the reference. Given the  lack of coverage of higher metal ions, it is not possible to constrain the 
ionization parameter in the absorbing gas. However, we note that 6 of the 9 sight lines  
with detected H I and metal lines have a DLA or a sub-DLA with log $N_{\rm H I } \ga$ 19.7, and are thus 
not expected to have significant ionization corrections. 

\subsection{Other Mg II Absorption Systems}

In addition to the { previously known Mg II absorbers mentioned in sec. 3.1}, weaker Mg II absorption systems were also detected in 
the available MagE/SDSS data. For Q1017-2046, a weak Mg II absorber at $z=0.7952$ with identical Mg II column densities of log N$_{\rm Mg II} = 12.38 \pm 0.06$, but no Fe II detection was detected in both sight  lines A and B. The HST STIS spectrum of Q1017-2046A shows no H I absorption at $z=0.7952$, but shows a weak absorption 
feature nearby. If this feature is H I Ly-$\alpha$ at $z=0.7987$, it may be associated with the Mg II absorber, { given that the redshift difference ($\approx 585$ km s$^{-1}$) is within 1 resolution element of STIS G230L. If this weak line is indeed H I Ly-$\alpha$, we can estimate a poorly constrained log $N_{\rm H I}$ value of $\approx 15.7 \pm 2.0$ assuming a Doppler $b$ parameter in the range of 10-30 km s$^{-1}$ for the $z=0.7987$ absorber toward Q1017-2046A. } However, the corresponding H I Ly-$\beta$ feature cannot be verified, because it lies in a region with very low quasar flux, owing to the Lyman limit break from the DLA at $z=1.086$.  The noisier HST STIS spectrum of Q1017-2046B shows no H I absorption at $z=0.7952$, or anywhere within 1500 km $s^{-1}$ of that redshift. { Based on the S/N in the STIS spectrum of Q1017-2046B, we can place a 3 $\sigma$ upper limit on the equivalent width of the  Ly-$\alpha$ line at $z=0.7952$ of $W < 0.91 $ {\AA}, which implies a 3 $\sigma$ upper limit on the H I column density of log $N_{\rm H I}$ $\le$ 18.19 in this sight line.}

For Q1349+1227, an Mg II absorber at $z=0.4913$ was 
detected in the combined (unresolved) SDSS spectrum with log $N_{\rm Mg II} = 13.92 \pm 0.20$, with marginal detections of Ti~II (log $N_{\rm Ti II} = 12.64 \pm 0.31$), and Fe II (log $N_{\rm Fe II} = 13.45 \pm 0.14$).  The corresponding H I Ly-$\alpha$ line lies in a very noisy part of the HST STIS spectrum (at wavelengths shorter than the Lyman-limit for the DLA at $z=1.2366$) and thus could not be measured in sight line A or B. 

Interestingly, we also note that no metal lines were detected at the estimated lens redshift of $z=0.23$ toward Q1054+2733 in the SDSS spectra. Likewise, no metal lines were detected at the estimated lens redshift of $z=0.645$ in 
the SDSS spectrum of Q1349+1227. { However, we note that the SDSS spectra are noisy, and we cannot rule out the presence of weak Mg II absorbers. These findings are consistent with other observations that also found that not all lenses produce strong absorption features (e.g., Rogerson \& Hall 2012; Zahedy et al. 2016). For example, Zahedy et al. (2016) found that 1 of the 3 lens systems in their study (the quadruple lens for HE 0405-1223, which arises in a group environment) shows no Mg II absorption  along any of the 4 sight lines. }

\section{Discussion}

{ We now discuss the implications of our results for comparing the H I and metal content between the individual sight lines in GLQ absorbers, keeping in mind that the derived column densities have the uncertainties discussed in sec.  2.6 and sec. 3.}

\subsection{H I Column Densities and their Differences between the Sight Lines}

For the $z=1.09$ absorber toward  Q1017-2046A and Q1017-2046B, both the Ly-$\alpha$ and Ly-$\beta$ features were used in estimating the H I column densities. The Ly-$\beta$ feature in the A sight line appears to be blended with another feature at a different redshift (possibly a Ly-$\alpha$ forest line at  a lower redshift.) The H I column densities derived are comparable along the two sight lines. 

The H I column densities along the two sight lines for the $z=0.68$ absorber toward Q1054+2733A,B  cannot be compared exactly, since we can only place an upper limit on the H I column density along the B sight line owing to the much noisier spectrum. The  Ly-$\beta$ line at $z=0.68$ is too noisy and blended in both sight lines  to give useful constraints. 
Additionally, we detect a strong feature in Q1054+2733A, { which may be H I Ly-$\alpha$ at $z=0.6687$, or Si III $\lambda$ 1206 at $z=0.6815$, or a blend. This feature is not seen in Q1054+2733B. This suggests that the cool gas in the $z=0.6687$ absorber and/or the warm gas in the $z=0.6815$ absorber detected toward Q1054+2733A has a transverse extent of $\lesssim$1.9 kpc.}

For the $z=1.24$ absorber toward Q1349+1227A and Q1349+1227B, we use both the Ly-$\alpha$ and Ly-$\beta$ lines in estimating the H I column densities. The Ly-$\beta$ feature for this absorber toward Q1349+1227B may seem over-predicted for the estimated $N_{\rm H I}$ value, but the fit is consistent with the data within the high noise level present near this feature {(see Fig. 4)}. If this noisy Ly-$\beta$ feature is excluded from the fit (i.e., if only the Ly-$\alpha$ feature is used), there is little change to the H I column density derived (log $N_{\rm H I}$ = 21.39$\pm$0.05 instead of 21.37$\pm$0.05.) In either case,  the best-fit H I column density appears to be higher along the B sight line than that along the A sight line by a factor of $\sim 3.0$.

For the $z=0.48$ absorber toward Q1355-2257A,B, the H I column density is higher along the B sight line by 0.62 dex (a factor of $\sim$ 4.2) compared to that in the A sight line. For the $z=0.70$ absorber toward Q1355-2257A,B, the H I column density is  higher along the B sight line by 0.92  dex (a 
factor of $\sim$ 8.3) compared to that in the A sight line. For both of these absorbers, only Ly-$\alpha$ lines were used in the fits, as the Ly-$\beta$ lines were not covered by the HST STIS data. 

Thus, there is a range in the variation of the H I column density across the few kpc distance between the two sight lines  in the GLQ fields studied here,  from 
almost no variation to variation by a factor of $\sim 8$. There is not a strong correlation between the difference in the H I column densities and the magnitude difference between the images; but this could be partly due to small number statistics. We further discuss our results together with others from the literature in section 4.3. 

\subsection{Element Abundances and Abundance Gradients}

We now describe our results for the Fe abundances [Fe/H] and the abundance gradients. We use 
[Fe/H] despite the fact that Fe is depleted on to dust grains simply because it is the most easily measured metal abundance { in our data}. We assume the abundance  gradient  to be uniform within each lens galaxy. We evaluate the [Fe/H] gradient from the difference in Fe abundances between the sight lines to the lensed quasar images and the difference in the impact parameters { (in kpc)} of the sight lines relative to the lens galaxy center, i.e., $\Delta$[Fe/H]/$\Delta$$r$ =  ([Fe/H]$_B$ - [Fe/H]$_A$)/($r_{B}$ - $r_{A}$). 

For the $z=1.09$ absorber toward Q1017-2046A,B (Figs. 7, 8), given that sight line B is at a smaller impact parameter from the galaxy center than sight line A, it is interesting that the Fe abundance along sight line B is somewhat lower than that along sight line A. The Fe 
abundances along the two sight lines  differ marginally { by 0.24 $\pm$ 0.15 dex}.  
Combining this with the impact parameters of the two sight lines ( 5.4   
kpc and  1.5    
kpc with respect to the galaxy center), we estimate an average abundance gradient of { +0.063 $\pm$  0.040} dex kpc$^{-1}$, i.e., a marginal inverted gradient compared 
to the negative abundance gradients seen in the Milky Way and nearby galaxies (e.g., $\sim$-0.01 to -0.09 dex kpc$^{-1}$ in the Milky Way, Friel et al. 2002; Luck \& Lambert 2011; Cheng et al. 2012;  -0.043 dex kpc$^{-1}$ in M101, Kennicutt et al. 2003; -0.027 $\pm$ 0.012 dex kpc$^{-1}$ in M33, Rosolowsky \& Simon 2008; -0.041$\pm$ 0.009 dex kpc$^{-1}$ in nearby isolated spirals, Rupke et al. 2010). 

For the $z=0.48$ absorber toward Q1355-2257A,B (Figs. 11, 12), the Fe abundance is lower along the B sight line { by 0.91 $\pm$ 0.33 dex} (a factor { of $\sim$6}). This is surprising if the lens galaxy is at $z=0.48$, given that sight line B passes closer to the galaxy center ( 1.6  kpc) than sight line A ( 5.7 kpc). In this case, we estimate an average abundance gradient { of +0.22 $\pm$ 0.08 dex kpc$^{-1}$}, which is stronger  and inverted compared to the abundance gradients seen in the Milky Way and nearby galaxies. 

For the $z=0.70$ absorber toward Q1355-2257A,B, metal lines are detected in sight line B (Fig. 13), but 
no metal lines are detected in sight line A. The non-detection of the strongest Fe II line covered (Fe II $\lambda$ 2383) implies a 3$\sigma$ upper limit on the Fe II column density of { log $N_{\rm Fe II} < 12.00$}, and hence an upper limit on the Fe abundance of { [Fe/H] $<$ -2.04}. The Fe abundance is higher along the B sight line { by $>$ 1.24 dex}  (i.e., a factor of $\gtrsim$17). 
If the lens galaxy is at $z=0.70$, the impact parameters of the A and B sight lines are 6.9  kpc and  1.9  kpc, respectively, implying an average abundance gradient { of $<$ -0.25 dex kpc$^{-1}$}. This abundance gradient is not inverted, but is much steeper compared to the abundance gradients seen in the Milky Way and nearby galaxies.  On the other hand, such a steep metallicity gradient is consistent with the strongly negative metallicity gradients found in some galaxies at $z \sim 1-2$ (e.g., Yuan et al. 2011; Jones et al. 2013). 

Given that the metallicity gradient in DLAs 
at $0.1 < z < 3.2$ is typically fairly weak (-0.022 $\pm$0.004 dex; Christensen et al. 
2014), one may wonder whether the steep abundance gradient in the $z=0.70$ absorber toward Q1355-2257 A,B is a result of 
differing degrees of Fe depletion. However, this would require sight line A to be more dusty, even though the H I column density in sight line A is smaller than that in sight line B by a factor of $\sim$11. (We note, however, that Chen et al. 2013 reported a differential X-ray absorption between images A and B, with image A having stronger X-ray absorption by $\Delta N_{H} = 4.0 \times 10^{20}$ cm$^{-2}$, which suggests the A sight line could be dustier.)

{Thus, if the redshift of the lensing galaxy for Q1355-2257A,B is $z=0.48$ as seems more likely to us (as explained in sec. 4.3), it has a positive abundance gradient. If, however, the less likely value of $z=0.70$ is in fact the correct redshift 
of the lens, it would have a negative abundance gradient. These different conclusions underscore the need for accurate determinations of the lens galaxy redshifts.}

Overall, our 
observations suggest a mixture of both positive (inverted) and negative metallicity gradients. The inverted gradients are interesting, since such gradients have been suggested as evidence of inflowing metal-poor gas at high redshift (e.g., Cresci et al. 2010; Queyrel et al. 2012). Indeed, Queyrel et al. (2012) have suggested that such inverted gradients may be found in metal-poor galaxies. We note, however, that the gradient in the dust-corrected metallicity could be different from the gradient in [Fe/H]. Since the dust 
depletion of Fe in DLAs correlates with metallicity (e.g., Kulkarni et al. 2015), the sight line with higher [Fe/H] is likely to have a more severe dust depletion and thus a higher intrinsic (dust-corrected) metallicity, implying a steeper gradient  (whether positive or negative). Indeed, in Q1017-2046A,B, the only case in our sample where we have measurements of Si (an element less depleted than Fe), the abundance of Si is higher than that of Fe by 1.13 $\pm$  0.15 dex in sight line A and by $0.73 \pm 0.24$ dex in sight line B. Furthermore, [Si/H] is lower in the B sight line by { 0.64} $\pm$ 0.24 dex, implying a [Si/H] gradient of { +0.167 $\pm$  0.063} dex, which is steeper (and even more inverted) compared to the [Fe/H] gradient of { +0.063 $\pm$  0.040}. { We also note  that the abundance gradients depend on the H I column densities, whose accuracy is limited by the low spectral resolution of the STIS G230L spectra, as discussed in sec. 2.6.1.} Other sources of uncertainties in our estimates of abundance gradients come from the fact that they are based on only two sight lines through the galaxy, and that they involve the assumption of a uniform gradient. 

Another way to estimate metallicity gradients in galaxies producing DLA/sub-DLAs is by comparison of emission and absorption metallicities: the emission typically comes from regions close to the galaxy center, while the absorption typically arises in regions further away. Integral field spectroscopic (IFS) observations allow the determination of emission-line metallicity (e.g., P\'eroux et al. 2011a, 2012, 2013). These studies also find a mixture of positive and negative metallicity gradients. Our results from GLQ sight lines  are thus consistent with the IFS results. We compare our results with 
the IFS results and other studies of abundance gradients in more detail in section 4.5. 

\subsection{Comparison with Other DLA/sub-DLA Absorbers toward GLQs}

We now compare the absorbers studied here with absorbers toward other GLQs, where at least one of the multiple sight lines shows a DLA or a sub-DLA. Only a {small number} of other GLQs (HE1104-1805,  HE 0512-3329,  Q0957+561, H1413+117, UM673, HE 0047-1756{, SDSS J1442+4055}) have measurements of H~I column densities in such absorbers  (Lopez et al. 1999; Lopez et al.  2005; Churchill et al. 2003; Monier et al. 2009;  Cooke et al. 2010; Zahedy et al. 2017; { and Krogager et al. 2018} respectively). { Table 26 summarizes the sample including our own measurements.}  
For HE1104-1805, HE 0512-3329, Q0957+561, UM673, HE 0047-1756, { and SDSS J1442+4055}, measurements of metal line column densities also exist in the same absorbers. 
We therefore include these absorbers along with the absorbers studied here in our study of abundance variations in absorbers toward GLQs (those arising in the lens galaxies as well as other absorbers). 

In the case of HE 1104-1805, Q0957+561, UM 673, { and SDSS J1442+4055}, the DLA/sub-DLAs reported in the literature are at redshifts considerably higher than the lens galaxy redshifts. For HE 0512-3329, while the  lens redshift is not certain, the lens is likely to be the galaxy associated with the DLA at $z=0.93$ (Gregg et al. 2000). Indeed, a DLA is detected along the sight lines of both HE 0512-3329A and HE 0512-3329B, separated by 0.64\arcsec, with essentially identical H I column densities (log $N_{\rm H I} = 20.49 \pm 0.08$ and log $N_{\rm H I} = 20.47 \pm 0.08$, respectively, and this is likely to be the lensing galaxy (Lopez et al. 2005). Furthermore, Lopez et al. (2005) found some evidence of difference in metallicities ([Fe/H] = $-1.52 \pm 0.11$, [Mn/H] = $-1.44 \pm 0.09$ in sight line A and [Fe/H]  $>-1.32$ dex, [Mn/H] = $-0.98 \pm 0.09$ in sight line B.  For HE 0047-1756, a sub-DLA has been detected by Zahedy et al. (2017) at the redshift of the lens galaxy in both sight lines, separated by 1.44\arcsec, with similar H I column densities (log $N_{\rm H I} = 19.7 \pm 0.1$ and log $N_{\rm H I} = 19.6^{+0.2}_{-0.3}$, respectively, for sight lines A and B). Fe II lines were also detected along both sight lines, implying abundances [Fe/H] of $-0.59 \pm 0.10$ for HE 0047-1756A and $-0.40 \pm 0.26$ for HE 0047-1756B. We therefore include the absorbers toward HE 0512-3329 and HE 0047-1756 along with the absorbers toward Q1017-2046 and Q1355-2257 in our examination of lens galaxy properties. Thus, our data { (which provide H I and metal column densities for 2 lens galaxies)} have doubled the sample of lens galaxies with measurements of H~I column densities and element abundances along multiple sight lines. For the sake of definiteness, we assume the redshift of the lensing galaxy for Q1355-2257 to be 0.48 in the discussion below (but we also consider the alternative redshift of 0.70 in sections 4.2 and 4.4). 

We now describe the differences between the absorber properties along the two sight lines for each of the above-mentioned GLQs and search for any trends in these differences with sight line separations.

\subsubsection{Transverse Separations} 

{ In a lensing geometry the light paths of lensed images at redshifts beyond the lens converge to the source, and therefore the intrinsic separation between the images gets smaller toward the source. When the absorber reshift is greater than or equal to the lens redshift,} 
the transverse separation $l_{AB, a}$ between the sight lines at the absorber redshift can be calculated as 

\begin{equation}
l_{AB, a} = {D_{aq} (1+z_{l}) \Delta \theta_{AB} D_{l} \over {D_{lq} (1+z_{a})}}, 
\end{equation}
where $D_{l}$, $D_{lq}$, and $D_{aq}$ denote the angular diameter distances between the observer and the lens, the lens and the quasar, and the absorber and the quasar, respectively, and $ \Delta\theta_{AB}$ is the angular separation between quasar images A and B. It is worth noting that $D_{aq} \ne D_{q} - D_{a}$, $D_{lq}  \ne D_{q} - D_{l}$,  $D_{aq} \ne -D_{qa}$, and $D_{lq} \ne -D_{ql}$. Following Hogg (1999), we use 

\begin{equation}
D_{12} = {c \over {H_{0} (1+z_{2})}}  \int_{z_{1}}^{z_{2}} [\Omega_{\Lambda} + \Omega_{m}(1+z)^{3}]^{-1/2} \, dz. 
\end{equation}
In the special case when the absorber is the lens itself, $l_{AB, a} = \Delta \theta_{AB} D_{l}$. In the more typical case of $z_{a} > z_{l}$, the separation at the absorber redshift is smaller than that at the lens. Indeed, the closer the absorber is to the quasar, the smaller the separation probed at the absorber redshift. 

\subsubsection{{ Trends with Sight Line Separation and Quasar Image Brightness}}

We now attempt to examine whether (a) the absolute differences in H I column densities, Fe II column densities and Fe abundances depend on the separation between the sight lines, and (b) whether these properties along the sight line to the fainter image are systematically smaller (or larger) compared to those along the sight line to the brighter image. { We note that for non-lens absorbers, if the absorbing galaxy has not been imaged, it is not known how the sight line separation 
relates to the impact parameters, since this relation depends on the location of the galaxy center relative to the two sight lines. Nevertheless, even in such cases, it is of interest to examine whether and how absorber properties correlate with the sight line separation. Indeed, several past studies have examined whether equivalent width differences between the sight lines are correlated with the separations (e.g., Rauch et al. 2001; Ellison et al. 2004). For example, if the sight line separation is smaller than the absorber size, then one may expect that smaller the separation, the smaller should be the column density difference.} At large separations the trend may disappear as the separation exceeds the absorber size. Furthermore, one may expect the fainter sight line to have higher H I  column density and metallicity if this sight line is obscured more due to a dustier absorbing region (although the lensing configuration also influences the relative brightness of the images). 

Fig. 15a shows the absolute difference in the H I column densities   $\lvert$$\Delta$ log $N_{\rm H I}$$\rvert$ vs. the { transverse} separation between the sight lines at the absorber redshift $l_{AB, a}$ { calculated using equations (1) and (2)}. { A number of the points in Fig. 15a are for absorbers along the sight lines toward a quadruple lens, and thus may not be independent of each other. Nevertheless, even if all the measurements are 
considered independent,} no significant correlation is observed between  $\lvert$$\Delta$ log $N_{\rm H I}$$\rvert$ and the sight line separation; we find a Spearman rank order correlation coefficient  of $r_{S}$ = 0.0726, and a 1-tailed  probability of  0.372 for this occurring by chance. While some absorbers show 
$>1$ order of magnitude difference in H~I column densities between the sight lines  over separations $>$3 kpc, others 
show smaller differences (including some with essentially zero difference) for separations of 6-8 kpc. This suggests 
a wide variety of sizes in the H I extent of DLA/sub-DLAs in the entire range studied (0-8 kpc). It is possible that the sight lines are being probed by different galaxies; such  a conclusion would be consistent with  a similar conclusion by Ellison et al. (2007) based on observation of coincident DLAs on a scale of 100 kpc toward a binary quasar.  Alternately, it may suggest a clumpy medium sampled randomly by a few sight line pairs. 

Fig. 15b shows the absolute difference in the Fe II column densities  $\lvert$$\Delta$ log $N_{\rm Fe II}$$\rvert$ between the sight lines vs. the separation between the sight lines at the absorber redshift { $l_{AB, a}$}. Not much correlation is seen here either, with a Spearman rank order correlation coefficient of { -0.0476}. Fig. 15c shows the absolute difference in the Fe abundances $\lvert$$\Delta$[Fe/H]$\rvert$ between the sight lines vs. the separation between the sight lines at the absorber redshift.  Once again, no significant correlation is evident, with a Spearman rank-order correlation coefficient of -0.156. 

To examine differences between the sight lines toward the fainter and brighter images and the potential dependence of any such differences on the sight line separations, we now plot the actual differences (not their absolute values), i.e.,   the $N_{\rm H I}$, $N_{\rm Fe II}$, and [Fe/H] values along the B (fainter) sight lines minus the corresponding quantities in the A (brighter) sight lines vs. the separations. 
(We note that if we plotted these differences as ``A-B'' differences instead of ``B-A'' differences, any trends seen with sight line separation would be reversed; but this would have no impact on any conclusions reached regarding the relative differences between the sight lines to the brighter images and the fainter images.) 

Fig. 16a shows the difference in the H I column densities   $\Delta$ log $N_{\rm H I}$ = log $N_{\rm H I}^{B}$ - log $N_{\rm H I}^{A} $ vs. the transverse separation between the sight lines at the absorber redshift $l_{AB, a}$. 
Once again, { if all the measurements are 
considered independent,} over the range of separations covered, there is not much 
correlation between $\Delta {\rm log} N_{\rm H I}$ and  $l_{AB, a}$, with a Spearman rank order correlation coefficient { of $r_{S}$ =  -0.0148}, and a 1-tailed  probability  of  0.472 for this occurring by chance. { For 15 out of 23 cases,}  $\Delta$ log $N_{\rm H I} < 0$, i.e., the H~I column density is lower along the fainter sight line compared to that along the brighter sight line. { However, we caution that { 12 of the 23} measurements { are for two absorbers (6 points per absorber)} along the sight lines toward a quadruple lens{, and thus} may not be independent of each other. { If these 12 measurements are excluded, the fraction of cases showing $\Delta {\rm log} N_{\rm H I} < 0$ becomes smaller (6 out of 11), and the lack of correlation between $\Delta {\rm log} N_{\rm H I}$ and  $l_{AB, a}$ persists, with  { $r_{S}$ =   0.109}, and a 1-tailed  probability of  0.374 for this occurring by chance. }

Fig. 16b shows the difference in the Fe II column densities between the sight lines vs. the separation between the sight lines at the absorber redshift. A wide range of differences in Fe II column densities are observed, from -2.3 dex to $>$1.8 dex. In general, the sight lines with larger Fe II differences are those with larger H I differences. There is not much correlation between the difference in Fe~II column densities and the transverse separation, with a Spearman rank order correlation coefficient of  $r_{S}$ = { -0.119}. 

 Fig. 16c shows the difference in the Fe abundances $\Delta$[Fe/H] between the sight lines vs. the separation between the sight lines at the absorber redshift. While the sample is still small, there is a slight excess of cases where $\Delta$[Fe/H] $> 0$, i.e., [Fe/H] is larger along the fainter sight line than along the brighter sight line. There is { not much correlation} between $\Delta$[Fe/H] and the sight line separation, with a Spearman rank-order correlation coefficient of {  -0.524}. 

We note that the observation of lower $N_{\rm H I}$ values in fainter images for some GLQs may suggest that the faintness of the fainter images relative to the brighter images is not always caused by stronger obscuration in a more gas-rich and dust-rich region along the sight line. In other words, obscuration along the sight line may play a smaller role than the lensing configuration in determining the relative image brightnesses. { (We note, however, that the excess of lower-$N_{\rm H I}$ systems seen in the present sample is driven by systems without metal abundance measurements. Measurements of abundances of volatile and refractory elements in these absorbers are essential to determine how the dust depletions in these lower $N_{\rm H I}$ absorbers compare with those in the higher $N_{\rm H I}$ absorbers.) { Furthermore, a number of the available H I measurements are for { two} absorbers along sight lines toward a 
quadruple lens, and therefore may not be independent of each other. Expanding the sample to a larger number of independent GLQ sightlines is essential to determine whether any statistically significant trend is present between H I column density difference and sight line separation.}}

 \subsection{Comparison with Other Abundance Gradient Measurements} 
 
 We now examine the abundance gradients in GLQ absorbers at the lens redshifts with measurements of abundance gradients in other 
 galaxies. In particular, we examine whether the abundance gradients show any trends with the abundances at the galaxy centers.  The blue circles in Fig. 17 show, for the GLQ absorbers at the lens redshifts, the [Fe/H] gradient in dex kpc$^{-1}$ between the GLQ sight lines 
 vs. the  Fe abundance at the galaxy center estimated from the two measurements, assuming a uniform abundance gradient within the galaxy, i.e. [Fe/H]$_{0}$ = [Fe/H]$_{A}$ - $r_{A} \Delta$[Fe/H]/$\Delta$ $r$. { Some GLQ absorbers appear to have positive (i.e., inverted) gradients, although the uncertainties are substantial.}
  
 For comparison, the green stars in Fig. 17 show measurements of the abundance gradients inferred from long-slit spectroscopy or integral field spectroscopy (IFS) of absorption-selected galaxies at 0 $<$ $z$ $<$ 2.4 (primarily at 0.5 $\la$ $z$ $\la$1) foreground to quasars (Schulte-Ladbeck  et al. 2004; Chen et al. 2005; P\'eroux et al. 2012, 2016, 2017; Rahmani et al. 2016; and references therein), plotted vs. the central emission-line abundance in the galaxy. We refer to these latter measurements as ``QGP'' (quasar-galaxy pair) measurements.  For all QGPs, the oxygen abundance estimated from nebular emission lines for the galaxy is adopted as the central abundance. 
 The majority of these QGP emission-line metallicities (from P\'eroux et al. 2012; Krogager et al. 2013; Kacprzak et al. 2014; many from Rahmani et al. 2016) are based on [N II]/H-$\alpha$ flux ratios and the calibration of Pettini \& Pagel (2004; hereafter PP04). The measurement of Schulte-Ladbeck et al. 2004) is also based on [N II]/H-$\alpha$, but uses the calibration of Denicolo et al. (2002; hereafter D02). The measurements from Chen et al. (2005), Straka et al. (2016), some from Rahmani et al. (2016), and P\'eroux et al. (2017) are based on the $R_{23}$ index using the calibrations from Kobulnicky et al. (1999), who stated the 12 + log (O/H) vs. $R_{23}$ relation of Zaritsky et al. (1994; hereafter Z94) and parameterized the relation for 12 + log (O/H) as a function of  $R_{23}$ and log([O III]/[O II]) given in McGaugh et al. (1991; hereafter M91). The measurements from Fynbo et al. (2011, 2013) and P\'eroux et al. (2014) were based on the $R_{23}$ index using the calibration from Kobulnicky \& Kewley (2004; hereafter KK04). 
 
 Given that the different nebular line indices and/or calibrations can give significantly different metallicities, the prescriptions of Kewley \& Ellison (2008) were used to convert the measurements based on D02, Z94, M91, and KK04 
 to the [N II]/H-$\alpha$ calibration of PP04 for consistency with the values from P\'eroux et al. (2012), Krogager et al. (2013), Kacprzak et al. (2014), and Rahmani et al. (2016). The  emission-line metallicity values from P\'eroux et al. (2012) were adjusted slightly to account for the slightly different solar abundance of O adopted by us [12 + log (O/H)$_{\odot}$ = 8.69 from Asplund et al. (2009), as opposed to 8.66 from Asplund et al. (2004) adopted by P\'eroux et al. (2012)]. For each QGP, the abundance gradient was  estimated by comparing the central emission-line abundance thus derived to that in the absorber along the sight line to the background quasar. The uncertainties in the abundance gradients were calculated including the uncertainties in the re-calibrated emission-line metallicities, and the uncertainties in the absorption-line 
 metallicities, but they do not include uncertainties in the impact parameters of the quasar sight lines relative to the galaxy centers, or the uncertainties in the fitting parameters for the relations  converting the above-mentioned  calibrations to the PP04 calibration from Kewley \& Ellison (2008), since these uncertainties are not available. 
  
Some of the QGP measurements show positive abundance gradients, although the majority of the QGP values appear to be nearly zero or negative. The magenta squares in Fig. 17 show measurements from emission-line spectroscopy of arcs in gravitationally lensed galaxies at 1.5 $<$ $z$ $<$ 2.4 (Yuan et al. 2011; Jones et al. 2013, 2015). The remaining symbols show examples of other measurements from the literature: the black unfilled squares show the measurements for star-forming galaxies at $0.8 < z < 2.3$ from Swinbank et al. (2012), and the red hatched squares show measurements for isolated spiral galaxies at $z \sim 0$ from Rupke et al. (2010). The Rupke et al. (2010) measurements were based on the [N II]/[O II] diagnostic from Kewley \& Dopita (2002), and were converted to the [N~II]/H-$\alpha$ diagnostic with the PP04 calibration using the prescription in Kewley \& Ellison (2008). The metallicity values from Swinbank et al. (2012) were based on [N~II]/H-$\alpha$ with the PP04 calibration; so no corrections were needed for those. 
 
 The GLQ points (based on Fe/H) lie at systematically lower abundances than the QGP points (based on O/H), but this is likely an effect of depletion of Fe on dust grains, with possibly some contribution from $\alpha$-enhancement of O relative to Fe. If the [Fe/H] values for GLQs were  corrected for these effects, they would move to the right in Fig. 17. As mentioned earlier, correction for dust depletion would also make the gradients steeper. For example, measurements of Si (an element less depleted than Fe) are  available for both sight lines for one of our GLQs (Q1017-2046), and would give an estimated central Si abundance [Si/H] of -0.82 $\pm$  0.36 dex and an abundance gradient of {+0.167} $\pm$ 0.063 dex kpc$^{-1}$. 
 
{ There may be} an anti-correlation between the central abundance and the abundance gradient in the combined sample. The Spearman rank-order correlation coefficient for all the points plotted in Fig. 17, treating the limits as detections, { is $r_{S} = -0.379$, which is $\sim$ 3 $\sigma$ significant  
 (a 1-tailed probability of 0.0039 of this correlation occurring purely by chance)}, if the lens redshift for Q1355-2257 is taken to be 0.48 rather than 0.70. This trend is consistent with a weak anti-correlation found by Queyrel et al. (2012) between the gradient and the global galaxy metallicity, and with the anti-correlation found by Belfiore et al. (2017) between the gradient and stellar mass (which in turn is correlated with the central metallicity). Indeed, the GLQ measurements extend to lower central metallicities a similar trend reported by Jones et al. (2013; see their Fig. 8). 
 If the lens redshift for Q1355-2257 is 0.70 rather than 0.48, the significance of the correlation becomes weaker  ($r_{S}$ =  -0.279, with a 1-tailed probability of  0.0274 of this occurring purely by chance). Larger GLQ and QGP samples are needed to further assess the reality of such a trend. 
If such a trend exists, it could be caused by the infall of metal-poor intergalactic gas into the centers of the galaxies (e.g., Cresci et al. 2010; Queyrel et al. 2012), or due to central dilution resulting from mergers (e.g., Scudder et al. 2012; Ellison et al. 2013{; Thorp et al. 2019}). 
 Of course, as mentioned earlier, our GLQ abundance gradient estimates are based on only 2 sight lines per galaxy, and are subject to assumptions about symmetry, dust depletion, etc.{, and also limited by the uncertainties in the lens galaxy redshifts and in the measurements of column densities.} Despite these caveats and the small size of the GLQ sample, our findings demonstrate the potential for GLQs to offer constraints on  
 metallicity gradients in distant galaxies. These constraints, if confirmed with larger samples (including GLQs with more than two lensed images), and with results from other techniques, may offer new insights into evolutionary processes such as gas accretion, star formation, and chemical enrichment in galaxies.

\subsection{Masses of the Lensing Galaxies}

One benefit of targeting gravitationally lensing galaxies is that analysis of the images of the background lensed object allows determination of the mass distribution in the lens. For early-type galaxies, the mass distribution is often assumed to 
be that of a singular isothermal sphere (SIS), given by $\rho \propto r^{-2}$ (e.g., Koopmans et al. 2009). In other words, the lens matter is assumed to behave as an ideal gas in thermal and hydrostatic equilibrium that is confined by a spherically symmetric gravitational potential. 
For the case of a galaxy lensing a quasar, the velocity dispersion of an SIS that would produce the observed quasar image separation $\Delta \theta$ is given by $\sigma_{\rm SIS}^2 = c^{2}D_{q} \Delta \theta/(8 \pi D_{lq})$, where $D_{q}$ and $D_{lq}$ denote
the angular diameter distances between { the observer and the quasar}, and between { the lens and the 
quasar}, respectively. For the lens galaxies toward Q1017-2046, Q1054+2733, Q1349+1227, and Q1355-2257, we estimate velocity dispersions of 190, 167, 318, and 194 km s$^{-1}$, respectively { (which are comparable to the velocity dispersion of 192 km s$^{-1}$ estimated  for HE 0512-3329 by Wucknitz et al. 2003)}. These relatively large values support the lensing to arise in early-type galaxies. The imaging of the lens galaxies reveals a Sersic index of 3.7 (Q1054+2733) and 1.7 (Q1349+1227; Kayo et al. 2010).  
 
 The mass of the lensing galaxy can be {estimated} from the astrometry of the quasar images relative to the lens galaxy, as 
 $M= -c^{2}\Delta \theta_{AG} \Delta \theta_{BG} D_{q}D_{l}/(4GD_{lq})$, where $\Delta \theta_{AG}$ and $\Delta \theta_{BG}$ (which have opposite signs) are the angular separations of quasar images A and B from the lens { (e.g., Schneider et al. 1992)}.  For the lens redshifts adopted by us (including $z = 0.48$ for the Q1355-2257 lens), given the observed angular separations of the quasar images from the lens galaxies, we estimate masses log $(M/M_{\odot})$ of 10.80, 10.84, 10.82 and 11.02, respectively, for the lenses toward Q1017-2046, Q1355-2257, HE 0512-3329, and HE 0047-1756. The value for Q1355-2257 is consistent with the estimate of $5.0 \times 10^{10} M_{\odot}$ within the Einstein radius obtained by Morgan et al. (2003) assuming a lens redshift of $z=0.55$.  If the redshift of the lens toward Q1355-2257 is taken to be 0.70 instead of 0.48, the corresponding lens mass would be log $(M/M_{\odot})$ = 11.07. The estimate for HE 0512-3329 is consistent with the estimate of the mass enclosed within the Einstein radius of $M_{E} = 6.8 \times 10^{10} M_{\odot}$ (Wucknitz et al. 2003).  { Finally, we note that the estimates of the masses of the lensing galaxies are subject to uncertainties in the redshifts of the lensing galaxies.} 

 \subsection{Constraining the Mass-Metallicity Relation for Lensing Galaxies}
 
Measurements of metal abundances in the lens galaxy (from spectroscopy of the background quasar) along with the mass of the lens galaxy (from astrometry of the lensed images) can, in principle, offer a powerful technique to constrain the mass-metallicity relation (MZR) for the lens galaxies. While the standard ways of estimating the MZR rely on stellar mass, the lensing galaxies allow determinations of the total mass. 

 Fig. 18 shows a plot of [Fe/H] vs. the mass of the lens galaxy.  Panel (a) shows the mean of the [Fe/H] values along the two sight lines, while panel (b) shows the [Fe/H] estimated to be at the galaxy center based on the measurements at the two sight lines, assuming a uniform gradient. The GLQ data points are the measurements for Q1017-2046, Q1355-2257, HE 0512-3329, and HE 0047-1756{\footnote{{ Using an SIS model would change the GLQ absorber points in Figs. 18a,b minimally (by 0.01-0.17 dex in log $M/M_{\odot}$) and would not change any of our conclusions.}}}. For Q1355-2257, we show both the possibilities for the lens redshift ($z$=0.48 and $z$=0.70). For comparison, the green stars in panel b show the dynamical masses (inferred from the maximum gas velocities) and emission-line metallicities (assumed to be the galaxy central metallicities) for absorber galaxies in quasar-galaxy pairs observed with integral field spectroscopy or long-slit spectroscopy (adopted from Chengalur \& Kanekar 2002; Schulte-Ladbeck et al. 2004; Chen et al. 2005; P\'eroux et al. 2011b, 2012, 2014, 2016, 2017, Bouch\'e et al. 2013, 2016; Fynbo et al. 2011, 2013; Krogager et al. 2013; Rahmani et al. 2016). The magenta squares in panel b show the corresponding measurements for gravitationally lensed galaxies at $1.5 < z < 2.4$ (Yuan et al. 2011; Jones et al. 2013). Despite the 
 small number of measurements, it is interesting to compare the data with the metallicity vs. total mass relation for early-type galaxies. To this end, { we show the dynamical mass vs. stellar metallicity relation for early-type SDSS galaxies from Galazzi et al. (2006),  as a violet dashed curve in both panels of Fig. 18. Also, for comparison, we show as an orange dashed-dotted curve the expected metallicity vs. total mass relation obtained by using} the stellar mass to total (dynamical) mass relation for early-type galaxies 
 log~M$_{*} = (0.93 \pm 0.02 )$ log M$_{dyn} + (0.09 \pm 0.20)$ from Ouellette et al. (2017), { and the polynomial fit to the stellar mass vs. 
  stellar metallicity relation for quiescent galaxies at $z=0.7$ from Galazzi et al. (2014). The black squares show 
 the measurements for individual early-type galaxies at $z < 0.05$ from X-ray measurements for hot gas (Babyk et al. 2018).} 
 
 The measurements for QGPs and gravitationally lensed galaxies in Fig. 18(b) lie relatively close to 
 the expected trend for early-type galaxies. However, the GLQ points lie below the trend. { Of course, the gas-phase measurements for GLQ absorbers could, in principle, differ from the stellar measurements; but it is interesting that  the 
 GLQ absorber measurements are lower than the gas-phase measurements for QGPs and gravitationally lensed galaxies as well.}  One possible reason is that Fe is depleted onto dust grains, so [Fe/H] is lower than [O/H]. In the Milky Way, Fe is depleted on interstellar dust grains by $>1$ to $>2$ orders of magnitude in environments ranging from warm ISM to cool ISM (e.g., Savage \& Sembach 1996; Jenkins 2009). Furthermore, O may be enhanced relative to Fe due to $\alpha$-enhancement.  Both of these effects together could substantially lower [Fe/H] compared to the true metallicity. Unfortunately, absorption-line measurements of elements that do not deplete much on interstellar dust grains (e.g., O, S, Zn) are currently not available for the lens galaxy absorbers toward any of the GLQs.  As mentioned earlier, 
 measurements of Si (which is more depleted than O, S, Zn, but less depleted than Fe) are available for both sight lines for one of our GLQs (Q1017-2046). These measurements give an estimated central [Si/H] of  -0.82 $\pm$ 0.36 dex and a mean [Si/H] of -0.24 $\pm$ 0.12 dex, which would put that galaxy somewhat closer to the expected trend for early-type galaxies. Some 
 Fe depletion (possibly $\sim$ 0.5-0.6 dex) has also been suggested for HE 0047-1756 by Zahedy et al. (2017). It is thus plausible that, after correction for dust depletion and $\alpha$-enhancement, the MZR for lens galaxies may be consistent with the expected trend for early-type galaxies and with the other measurements. 
  
As seen in Figs. 18(a) and 18(b), three of the observed lens galaxies have quite similar inferred masses (within 0.04 dex if the lens toward Q1355-2257 is at $z=0.48$). The average [Fe/H] values for these systems differ by about 0.2 dex, but are consistent within the uncertainties. The differences are larger among the [Fe/H] values inferred for the galaxy centers, but so are the 
uncertainties in those values. These large uncertainties can be traced in part to the linear extrapolation made from the observed abundances for the two sight lines to the galaxy centers assuming a uniform gradient. (This effect is especially large when the impact parameters for the two sight lines are relatively close, e.g. for HE 0047-1756.) Additionally, the dispersion seen among the observed values in both panels could also be an effect of different levels of dust depletion and/or $\alpha$-enhancement among the observed lens galaxies. Measurements of element abundances (especially for undepleted elements such as S, O, Zn) in many more GLQ (and QGP) absorbers based on high S/N spectra are essential to obtain tighter constraints on the mass-metallicity relation in absorption-selected galaxies. 

\section{Summary and Outlook for Future Work}

We have measured absorption lines of H I as well as metals in 5 absorbers along multiple sight lines in 4 GLQs. Our data { (which provide H I and metal column densities for 2 lens galaxies)} have doubled the sample of lens galaxies with measurements of H~I column densities and metal abundances along multiple sight lines. Combining our data with the literature, we find no strong correlation between the absolute value of the difference in the H I column densities, Fe II column densities, or Fe abundances and the separation between the sight lines at the absorber redshift for separations of 0-8 kpc. { Absorbers toward fainter GLQ images show lower $N_{\rm H I}$ compared to those toward brighter images in 15 out of 23 cases, { potentially} suggesting that dust obscuration along the sight line may play a smaller role than the lensing configuration
in determining the relative image brightnesses;} { however, a number of the available H I  measurements are for absorbers along sight lines toward a quadruple lens, and therefore may not be independent of each other. Expanding the sample to a larger number of independent GLQ sightlines is essential to determine whether any statistically significant trend is present between H I column density difference and sight line separation.}

The abundance gradients inferred from some GLQ { absorbers} are positive (i.e., inverted). Combining our results with measurements for quasar-galaxy pairs studied with integral field spectroscopy or long-slit spectroscopy, and other measurements of abundance gradients from the literature, we find a tentative anti-correlation between the abundance gradient and the central metallicity of the galaxy, with metal-poor galaxies showing more positive abundance gradients, and metal-rich galaxies showing negative abundance gradients. Combining the  total masses inferred from the astrometry of the quasar images relative to the lens galaxies with the mean [Fe/H] measurements from the quasar spectroscopy, we find that the lens galaxies lie below the MZR expected for early-type galaxies at $z=0.7$ { and the measurements for QGPs and gravitationally lensed galaxies}. This difference may be caused { in part} by dust depletion of Fe and possibly also $\alpha$-enhancement of O relative to Fe. 

Observations of more DLA/sub-DLAs toward other GLQs are essential to examine how common the trends suggested by our small sample are. It is especially necessary to cover GLQs with small separations in order to put further constraints on small-scale structure in the absorbing gas. Additionally, it is essential to obtain higher resolution and shorter-wavelength UV observations to cover the Lyman series lines in order to obtain accurate H I column densities. It is also essential to obtain abundances of elements that do not deplete much on dust grains (e.g., S, O, Zn). For lens galaxies at $z < 1$, the measurements of S, O, Zn will require observations with UV and blue-sensitive spectrographs. Combining such measurements of undepleted elements with measurements of depleted elements in the same sight lines  will allow a direct determination of differences in dust depletion between the sight lines. Combining these results with existing differential extinction measurements in GLQ sight lines (e.g., Falco et al. 1999; Motta et al. 2002) will offer fresh insights into the spatial variations in sizes and chemical composition of interstellar dust grains in galaxies. Finally, {as noted earlier, 
 some of the lens redshifts are uncertain, which limits estimates of their distances, transverse separations of the sight lines, abundance gradients etc. It is thus} crucial to determine accurate lens galaxy redshifts in all GLQ cases. Future IFS observations of GLQ fields should enable such measurements with high efficiency and provide additional constraints on the lens galaxy masses and kinematics.

\acknowledgments
{ We thank an anonymous referee for many helpful comments that have improved this manuscript.} 
VPK and FHC gratefully acknowledge support from NASA/ Space Telescope Science Institute (grant HST-GO-{ 13801.001-A)} and the South Carolina Space Grant Consortium, and additional support from NASA grants NNX14AG74G { and NNX17AJ126G}. SL was partially funded by UCh/VID project ENL18/18.
{\it Facilities:} \facility{HST (STIS)}, \facility{Magellan (MagE)}.

\clearpage

\begin{figure}
\includegraphics[scale=0.75]{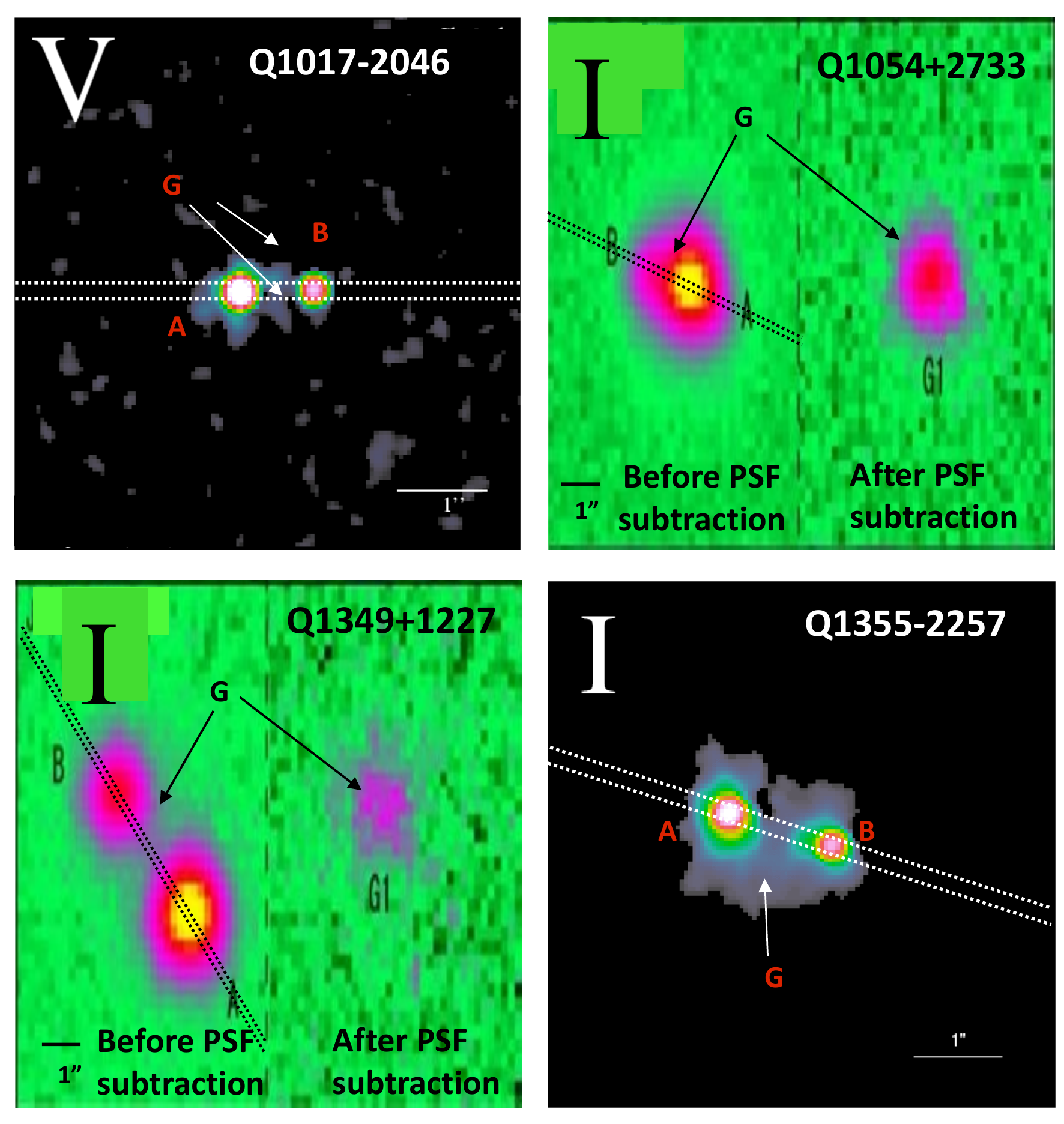}
\caption{HST or ground-based images of our targets in V or I band, showing the gravitationally 
lensed images of the background quasar and the foreground lens galaxy (adopted from the CASTLES gravitational lens  survey of Lehar et al. 2000 for Q1017-2046 and Q1355-2257, and from Kayo et al. 2010 for Q1054+2733 and Q1349+1227).  Dashed lines  show the STIS slit orientations used in our HST observations to 
efficiently observe the multiple sight-lines. For Q1054+2733 and Q1349+1227, the left half 
shows the quasar and faint lens galaxy images, while the right half shows the galaxy after 
subtracting the quasar images.}
\end{figure}

\begin{figure}
\includegraphics[width=15.3cm,height=6.0cm]{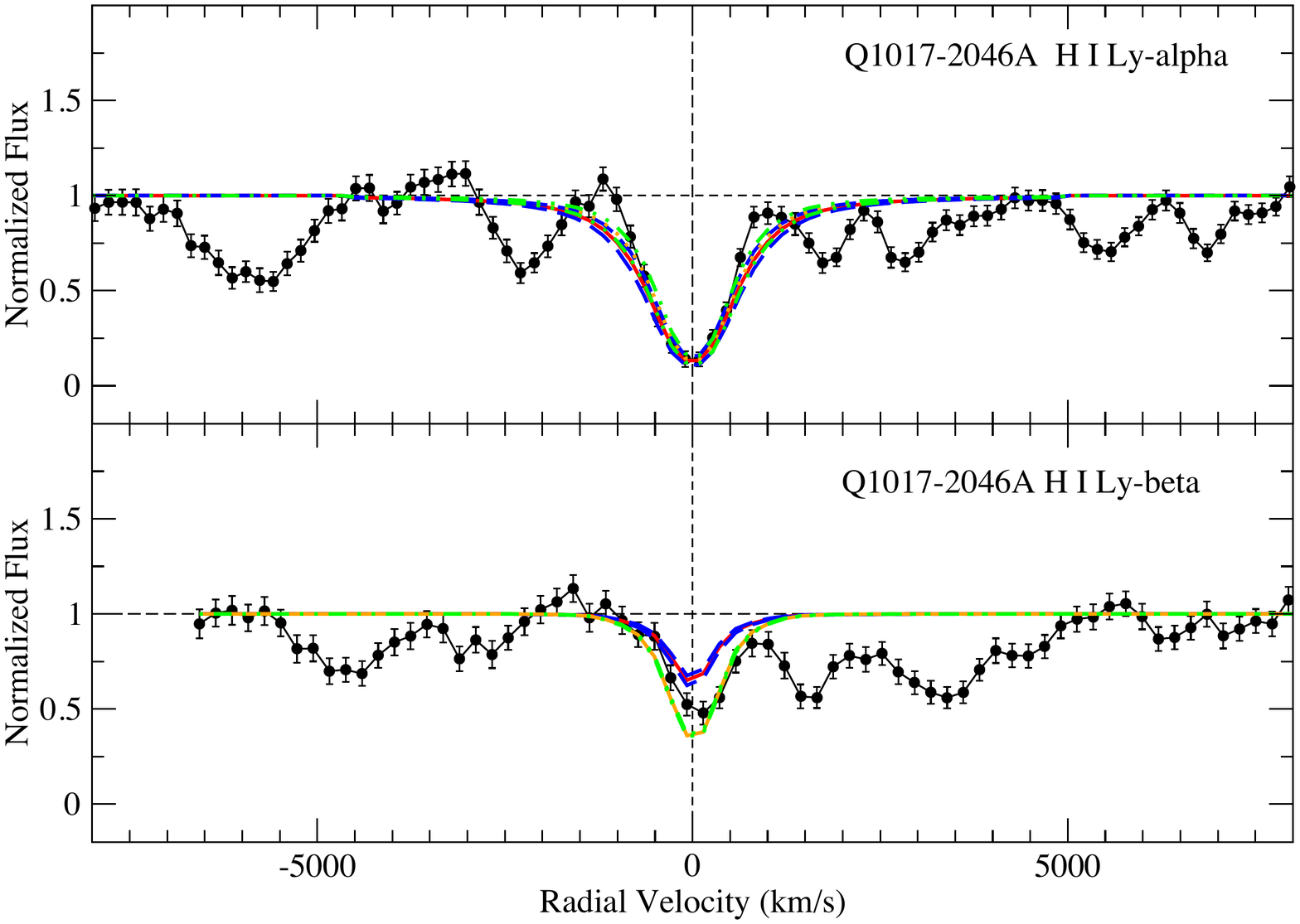}
\hskip 0.03in
\includegraphics[width=15.3cm,height=6.0cm]{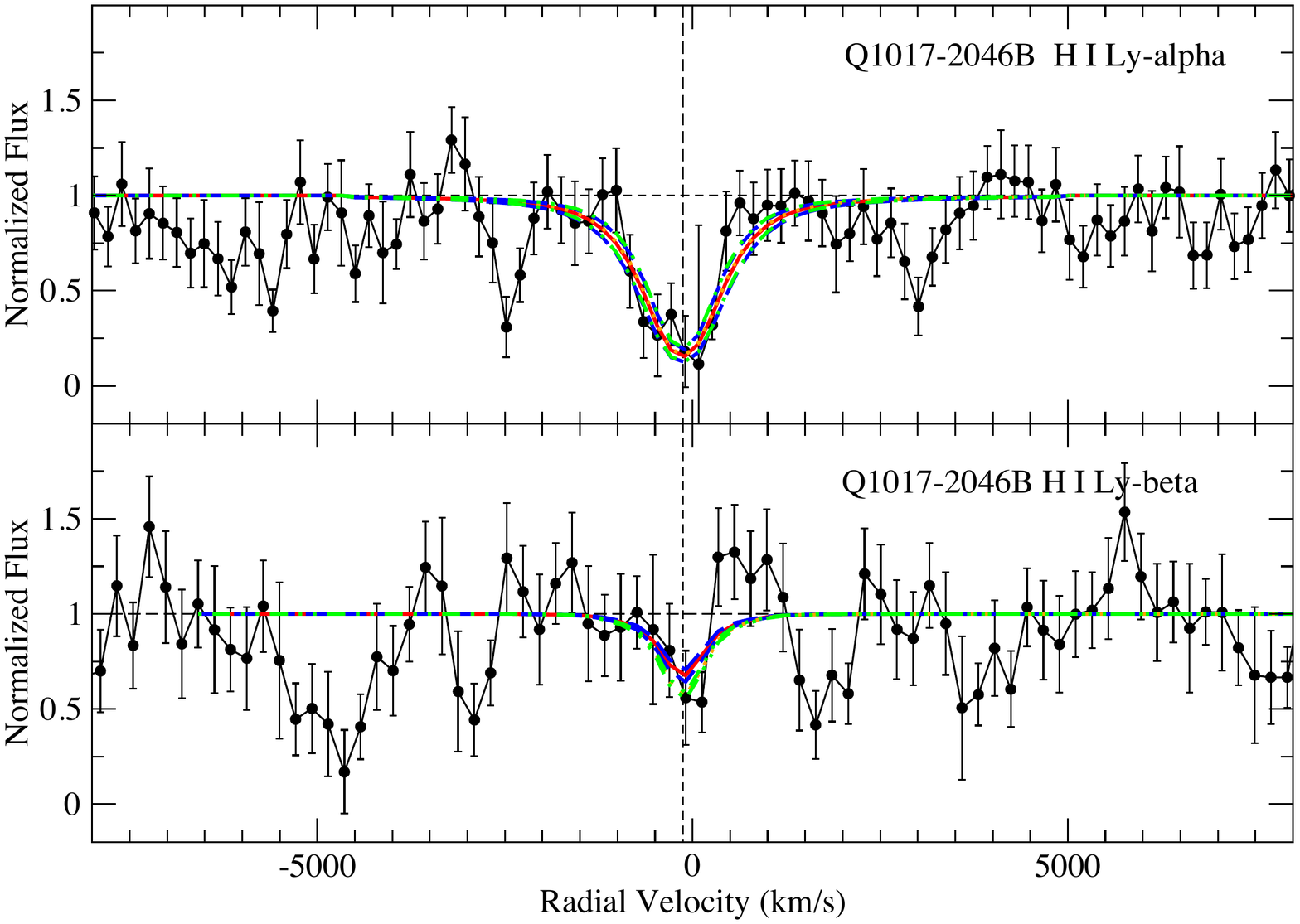}
\caption{HST STIS G230L spectra of Q1017-2046A (top two panels) and Q1017-2046B (bottom two panels) showing { velocity plots} for the region near Lyman-$\alpha$ and Lyman-$\beta$ absorption { with respect to $z = 1.08803$}. 
The curves show the Voigt profiles { (convolved with the instrumental profile)} obtained by fitting Ly-$\alpha$ and Ly-$\beta$ together. { Solid} red curves show the { fits for the minimum $b$ value considered, and dashed blue curves show profiles for the corresponding $\pm 1 \sigma$ uncertainties in $N_{\rm H I}$. Orange and green curves show profiles for the best-fit and $\pm 1 \sigma$ $N_{\rm H I}$ values, respectively, for the maximum $b$ value considered,  in case the $N_{\rm H I}$ values for the minimum and maximum $b$ values  differ by $\ge$0.02 dex.}
 { The horizontal dashed lines indicate the continuum levels. The vertical dashed lines indicate the positions of the components used in the fits. }
}
\label{fig_1017_HI_Lyab}
\end{figure}

\begin{figure}
\includegraphics[width=15cm,height=7.5cm]{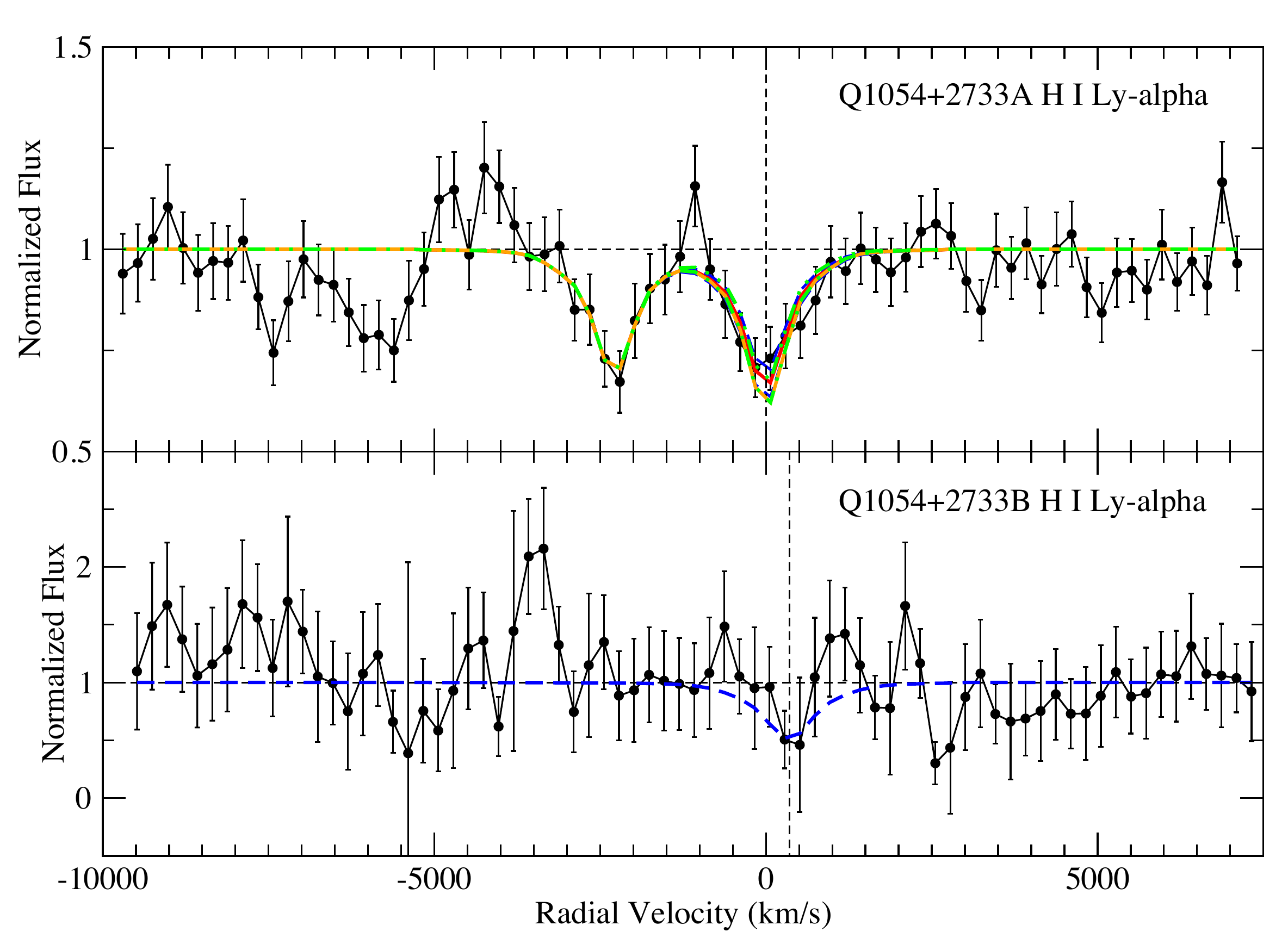}
\caption{HST STIS G230L spectra for Q1054+2733A,B showing {velocity plots for} the region near Lyman-$\alpha$ absorption with respect to $z =0.68153$. For Q1054+2733A, the red curve shows the best-fit Voigt profile for $b=15$ along with another absorption feature (which may be H I Ly-$\alpha$ at $z$ = 0.6687 or Si III $\lambda$ 1206 at $z$ = 0.6815, or a blend) not detected toward Q1054+2733B.  
The Ly-$\beta$ region is too noisy and blended to give useful constraints. The horizontal dashed lines indicate the continuum levels. The vertical dashed lines indicate the positions of the components used in the fits. For Q1054+2733A, the difference between the dashed blue curves and the red curve is that the former show Voigt profiles corresponding to the $\pm 1 \sigma$ uncertainties in log $N_{\rm H I}$ { for the minimum $b$ value considered} for the $z=0.6815$ absorber. { The orange and green curves show the  profiles for the best-fit and $\pm 1 \sigma$ $N_{\rm H I}$ values, respectively, for the maximum $b$ value considered.}
 For Q1054+2733B, only a dashed blue curve corresponding to the adopted upper limit on the H I column density is shown.  }
\label{fig_1054_HI_Lya}
\end{figure}

\begin{figure}
\includegraphics[width=15cm,height=6.5cm]{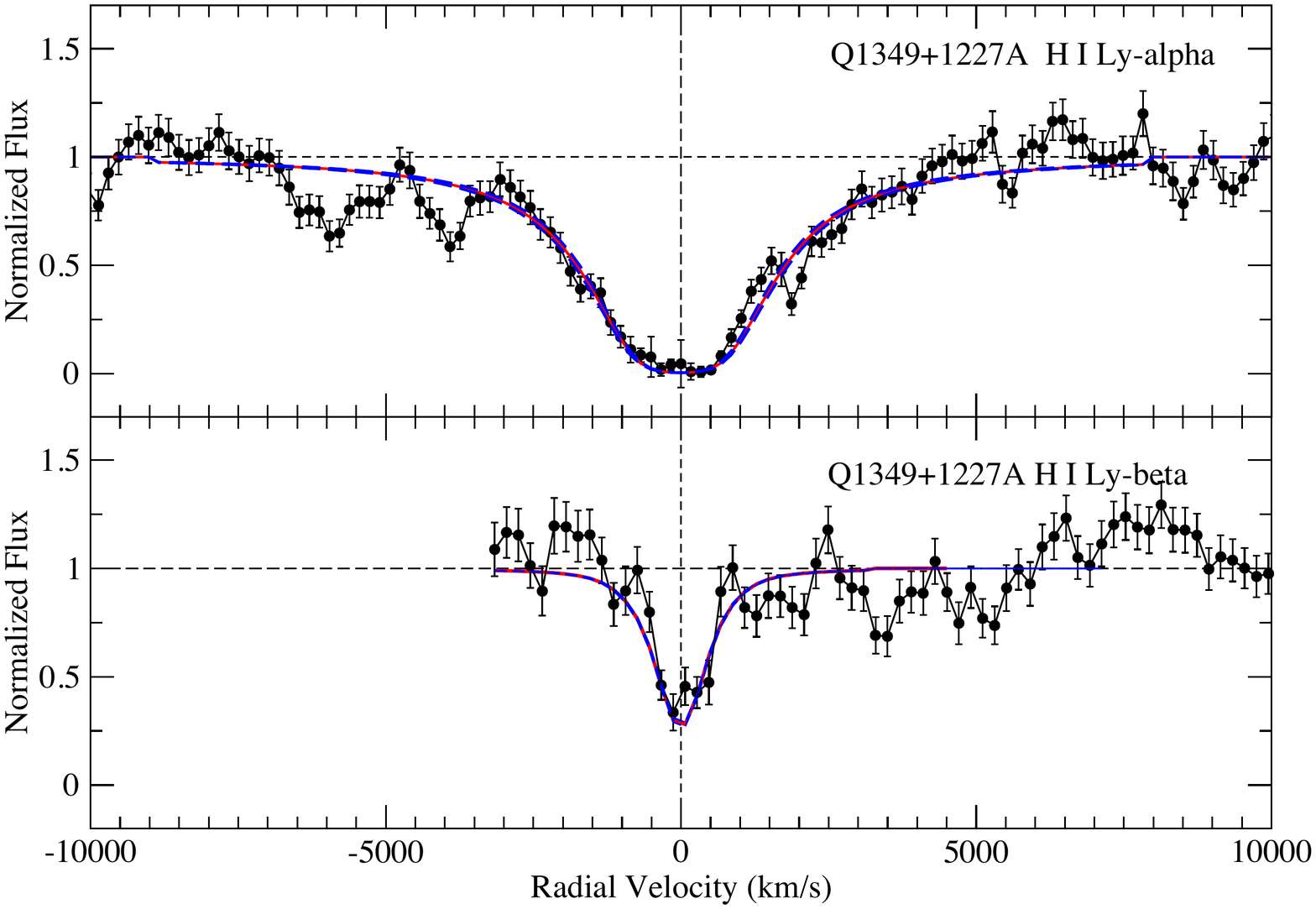}
\hskip 0.06in
\includegraphics[width=15.3cm,height=6.5cm]{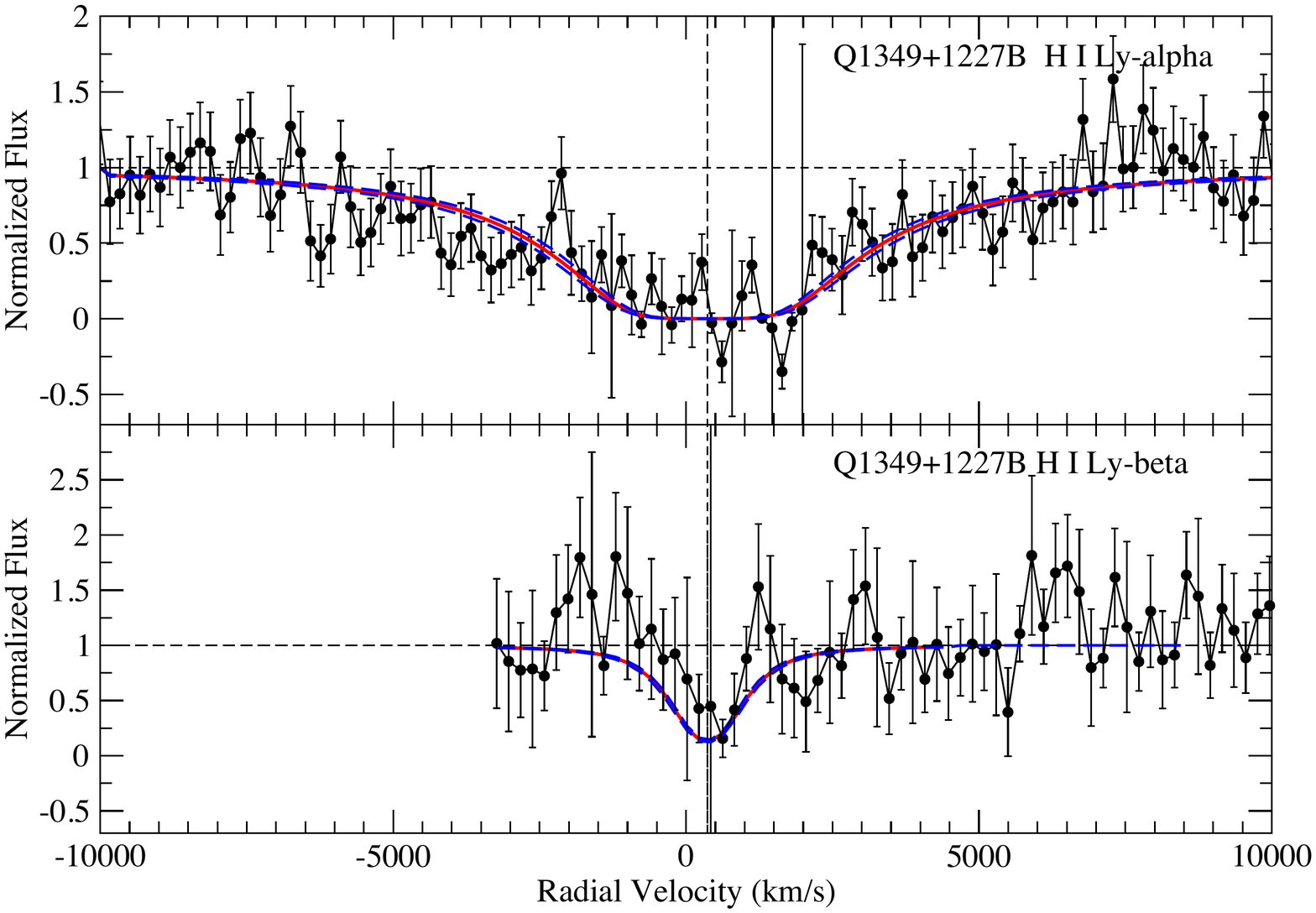}
\caption{HST STIS G230L spectra showing { velocity plots for} the region near Lyman-$\alpha$ and Ly-$\beta$ absorption for Q1349+1227A (top two panels) and Q1349+1227B (bottom two panels) { with respect to $z = 1.23750$}.  The red curves show the best-fit Voigt profiles obtained by fitting Ly-$\alpha$ and Ly-$\beta$ together { for the full range of $b$ values considered. (There is no difference between the $N_{\rm H I}$ values obtained for the minimum and maximum $b$ values.)  The dashed blue curves show Voigt profiles corresponding to the $\pm 1 \sigma$ uncertainties in log $N_{\rm H I}$.} The Ly-$\beta$ region for Q1349+1227B is very noisy, and ignoring it leads to very little change in the best-fit H I column density for this sight line. The horizontal dashed lines indicate the continuum levels. The vertical dashed lines indicate the positions of the components used in the fits. 
}
\label{fig_1349_HI_Lyab_fixN_20p86}
\end{figure}

\begin{figure}
\includegraphics[width=15cm,height=7.5cm]{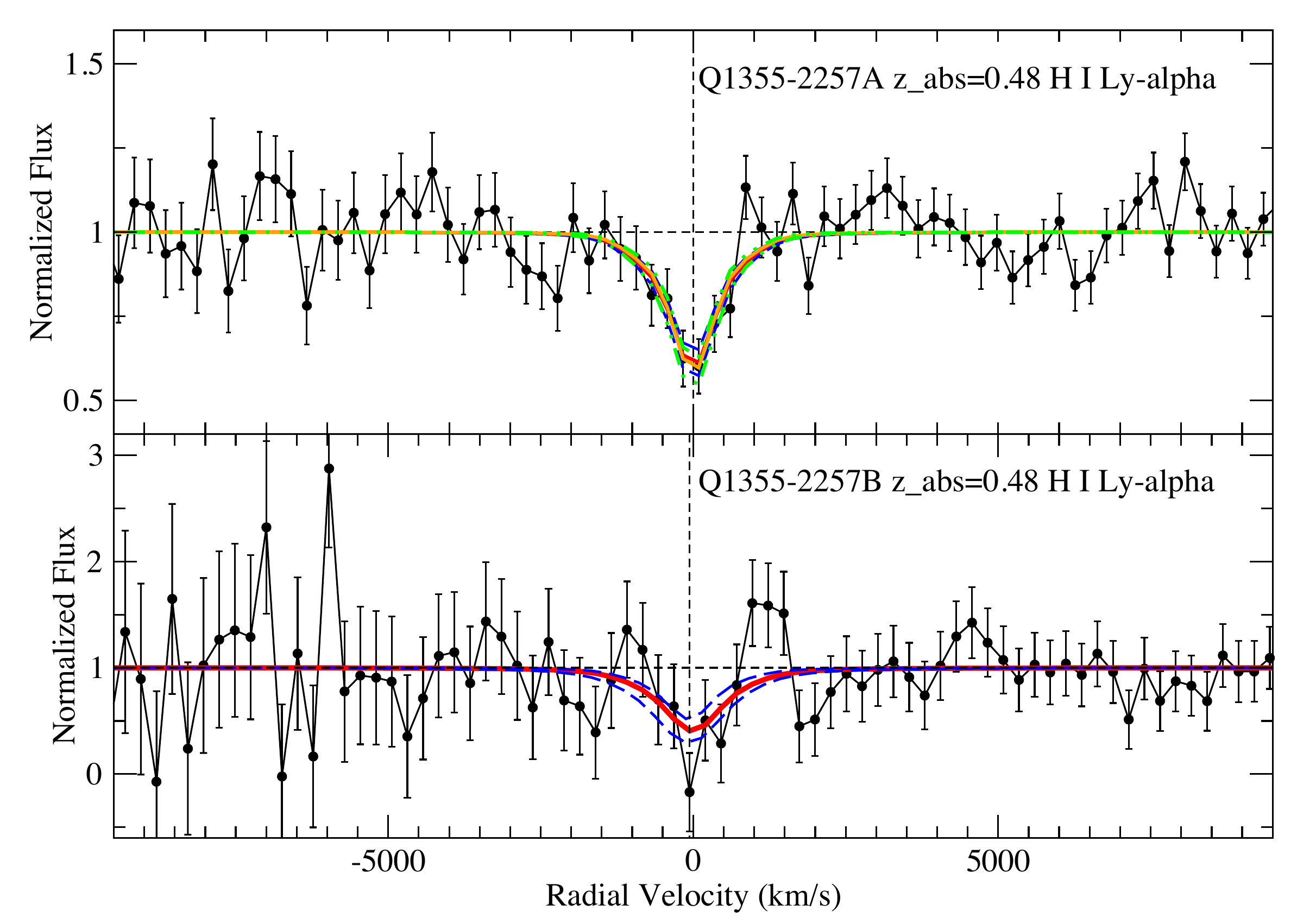}
\caption{HST STIS G230L spectra of Q1355-2257A,B showing velocity plots for the region near Lyman-$\alpha$ absorption  with respect to  $z = 0.48183$.  { Red curves show the best-fit Voigt profiles for the minimum $b$ value considered, 
and dashed blue curves show Voigt profiles corresponding to the $\pm 1 \sigma$ uncertainties in log $N_{\rm H I}$. Orange and green curves show the corresponding best-fit and $\pm 1 \sigma$ curves, respectively, for the maximum $b$ value considered, in case  the $N_{\rm H I}$ values for the minimum and maximum $b$ values  differ by $\ge$0.02 dex. }
 The horizontal dashed lines indicate the continuum levels. The vertical dashed lines indicate the positions of the components used in the fits.  }
\label{fig_1355_HI_Lyab}
\end{figure}

\begin{figure}
\includegraphics[width=15cm,height=7.5cm]{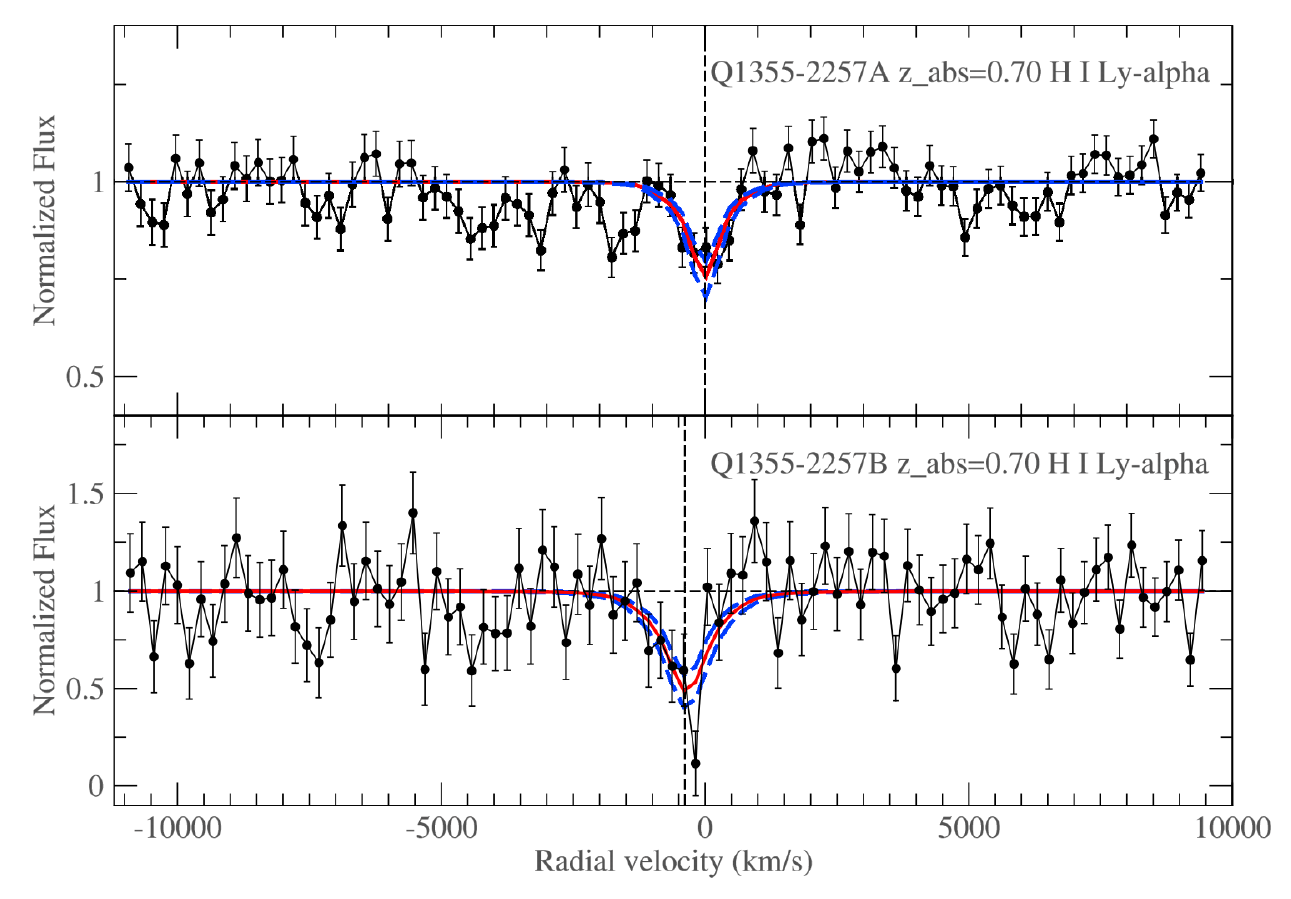}
\caption{HST STIS G230L spectra of Q1355-2257A,B showing velocity plots for the region near Lyman-$\alpha$ absorption with respect to $z=0.70721$.  { For both Q1355-2257A and Q1355-2257B,} the red curves show the best-fit Voigt profiles { for the minimum $b$ value considered}, and the dashed blue curves show Voigt profiles corresponding to the  $\pm 1 \sigma$ uncertainties in log $N_{\rm H I}$. { (No separate profiles are shown for the maximum $b$ value, since the $N_{\rm H I}$ values for the minimum and maximum $b$ value differ by $< 0.02$ dex for Q1355-2257B, and no $b$ value constraints are available for Q1355-2257A, where no metal lines are detected.)}  The horizontal dashed lines indicate the continuum levels. The vertical dashed lines indicate the positions of the components used in the fits. 
}
\label{fig_1355_z0p7_HI_Lyab}
\end{figure}

\begin{figure}[!th]
\includegraphics[scale=0.55]{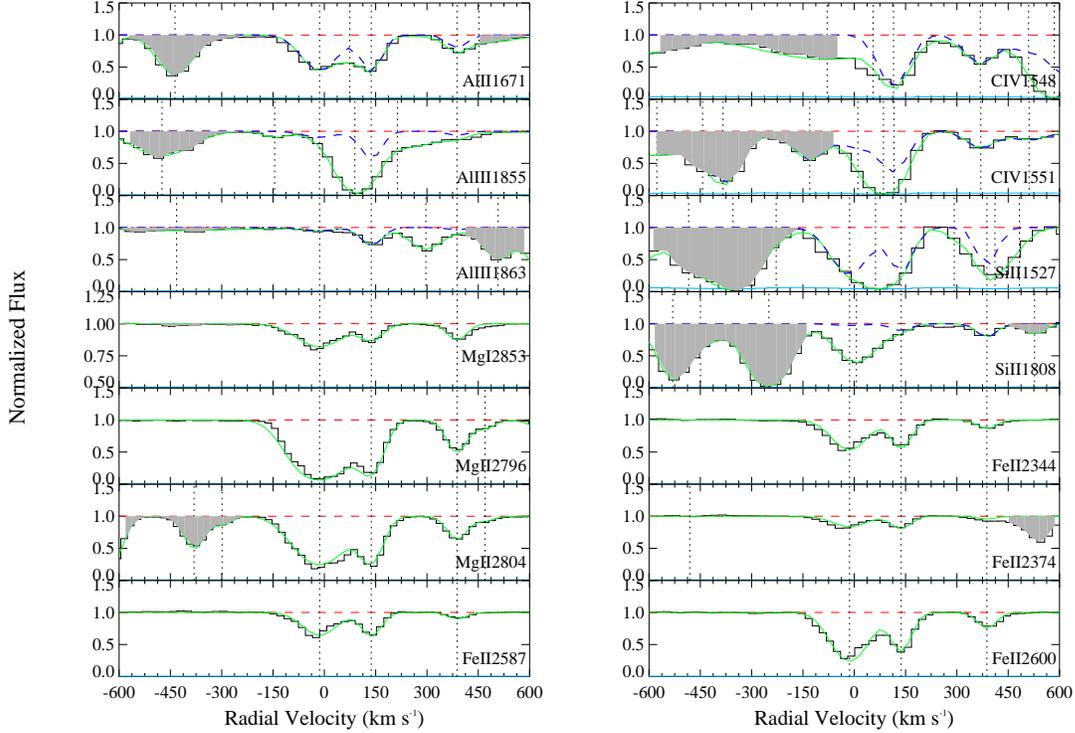}
\caption{ 
Velocity plots for the metal lines for the $z = 1.0859$ system in the spectrum of Q1017-2046A taken with the Magellan MagE spectrograph. In each panel, the normalized data are shown in black, the solid green curve indicates the theoretical Voigt profile fit to the absorption feature, and the dashed red line shows the continuum level. For lines in the Lyman alpha forest, the dashed blue curve indicates the profile fit of the metal line absorption feature alone. The $1\sigma$ error values in the normalized flux are represented by the { cyan} curves near the bottom of each panel. {Note that in a few panels with weak lines, if the normalized flux scale is shown starting at 0.5, the 1$\sigma$ error arrays are offset by 0.5, so that they can be viewed in the same panels.} The vertical dotted lines indicate the positions of the components that were used in the fit. Shaded regions indicate absorption unrelated to the line presented or regions of high noise. 
}
\label{fig_1017a_pro}
\end{figure}

\begin{figure}[!th]
\includegraphics[scale=0.6]{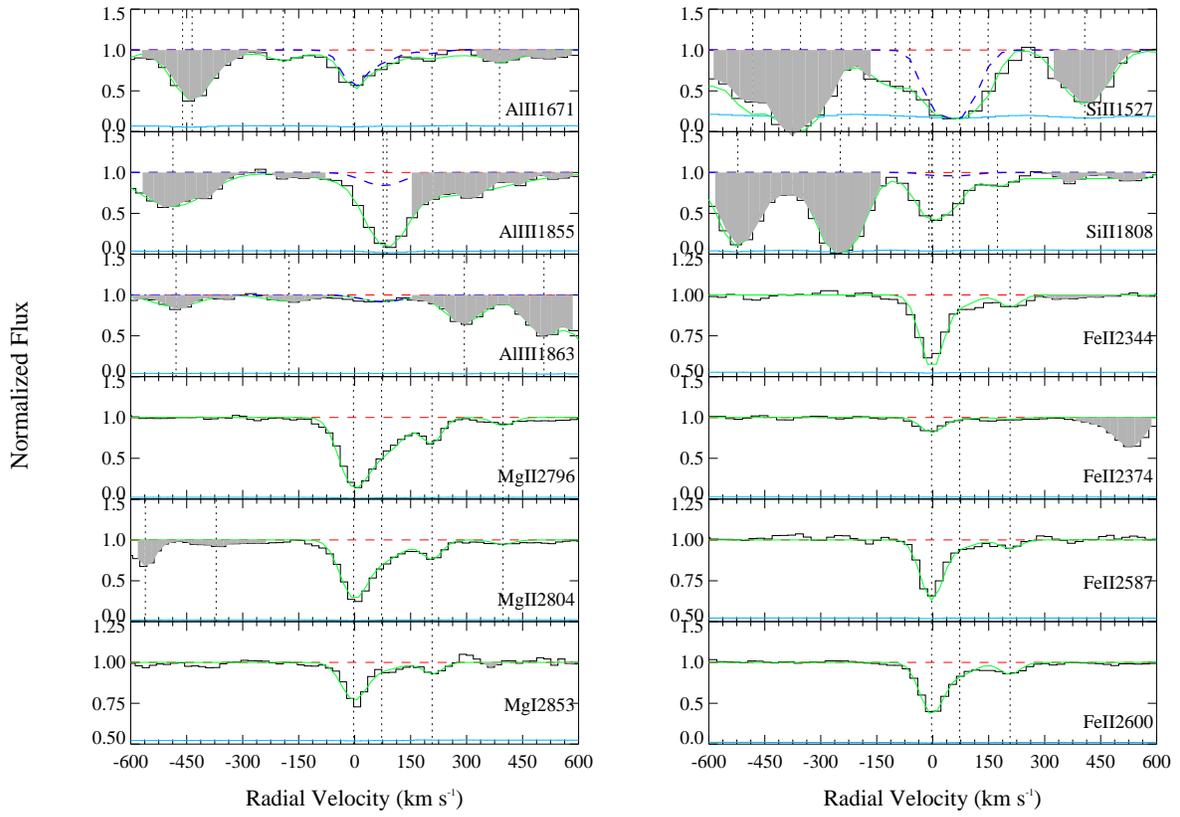}
\caption{ 
Same as Fig. 7, but for metal lines in the $z = 1.0859$ system toward Q1017-2046B.
}
\label{fig_1017b_pro}
\end{figure}

\begin{figure}[!th]
\includegraphics[scale=0.6]{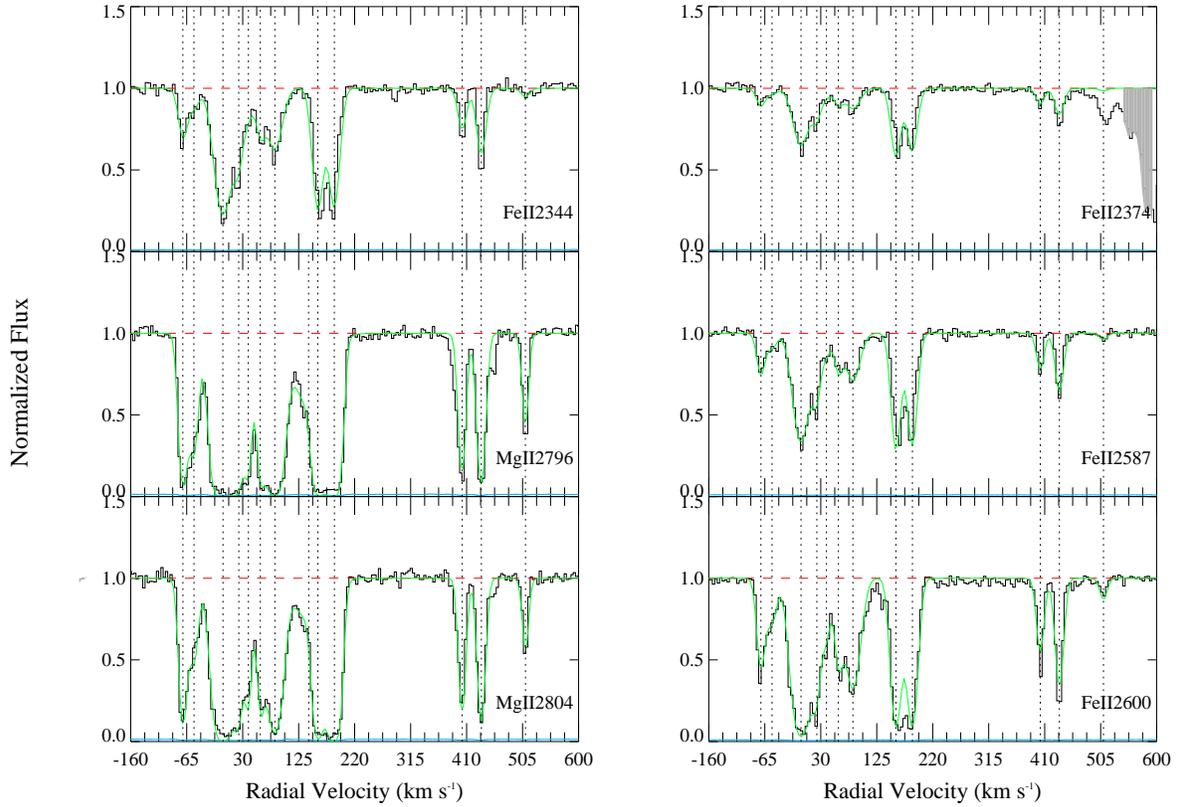}
\caption{ 
{Velocity plots for the metal lines for the $z = 1.0859$ system in the spatially unresolved archival  spectrum of Q1017-2046A+B obtained with the Keck HIRES spectrograph. In each panel, the normalized data are shown in black, the solid green curve indicates the theoretical Voigt profile fit to the absorption feature, and the dashed red line shows the continuum level. 
The $1\sigma$ error values in the normalized flux are represented by the cyan curves near the bottom of each panel. The vertical dotted lines indicate the positions of the components that were used in the fit. Shaded regions indicate absorption unrelated to the line presented or regions of high noise.}
}
\label{fig_1017ab_pro_hires}
\end{figure}

\begin{figure}[!th]
\includegraphics[scale=0.6]{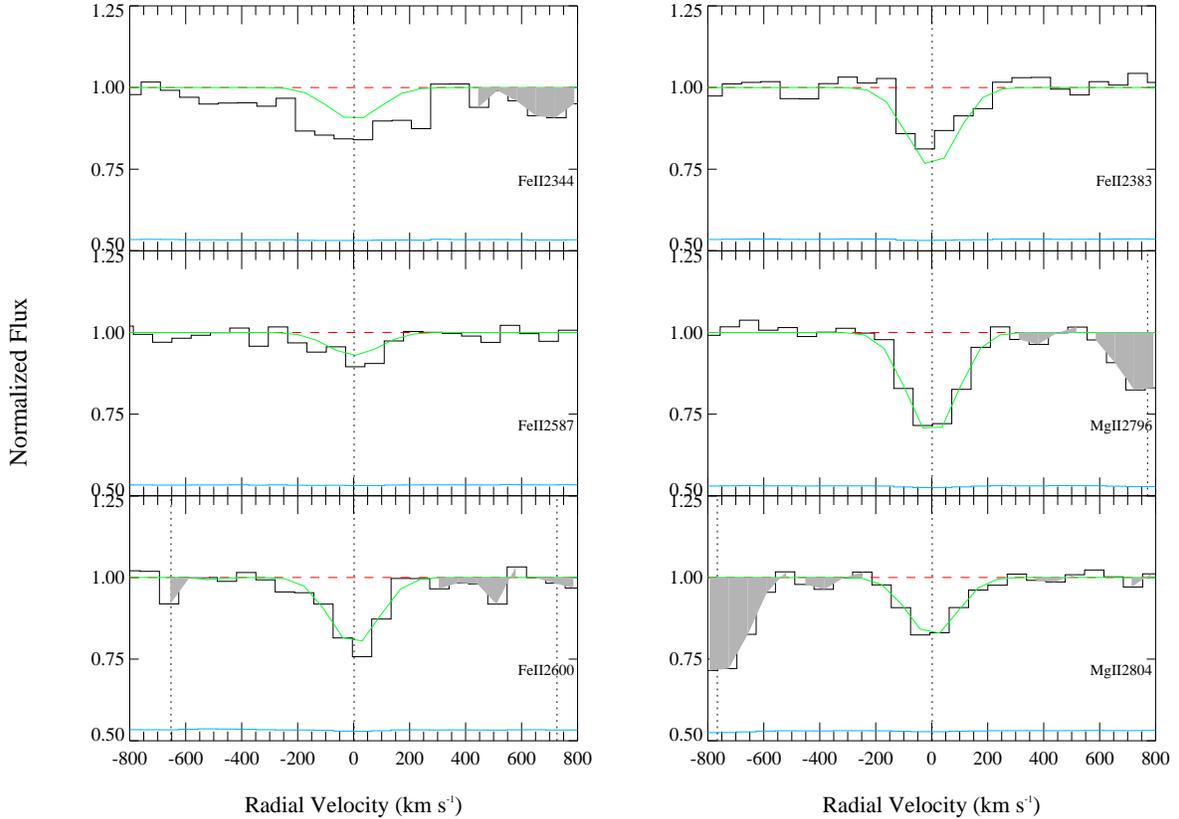}
\caption{ 
Velocity plots for the metal lines in the $z = 0.6794$ system in the SDSS spectrum of Q1054+2733. The two images of the quasar are not resolved at the SDSS resolution, and the spectrum shown is a superposition of the spectra along both sight lines. In each panel, the normalized data are shown in black, the solid green curve indicates the theoretical Voigt profile fit to the absorption feature, and the dashed red line shows the continuum level. The $1\sigma$ error values in the normalized flux are represented by the blue curves near the bottom of each panel. The vertical dotted lines indicate the positions of the components that were used in the fit. Shaded regions indicate absorption unrelated to the line presented.
}
\label{fig_1054a_pro}
\end{figure}

\begin{figure}[!th]
\includegraphics[scale=0.6]{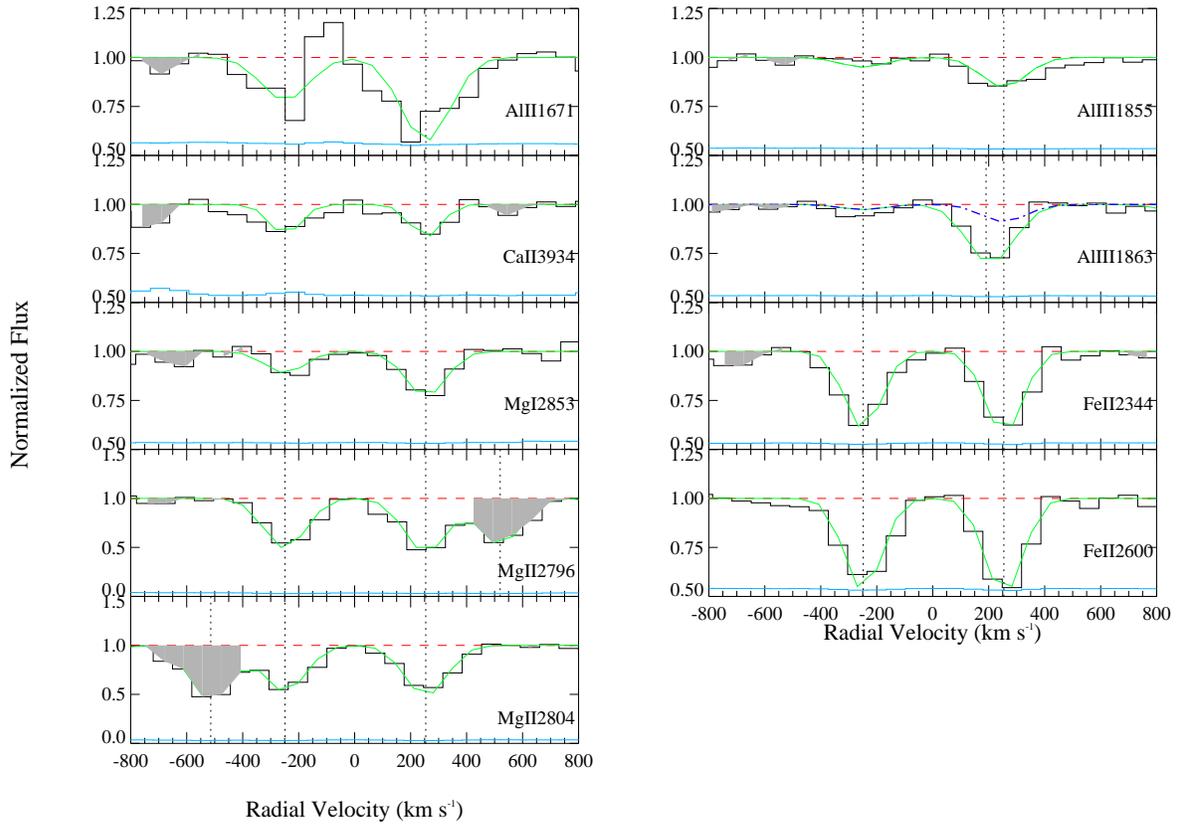}
\caption{ 
Same as Fig. 10, but for the $z = 1.2366$ system in the unresolved SDSS spectrum of Q1349+1227. The dot-dashed blue curve in the Al III 1863 panel shows the contribution of Mg II 2796 at $z=0.4913$ that is blended with the Al III 1863 line at z=1.2366. 
}
\label{fig_1349a_pro}
\end{figure}

\begin{figure}[!th]
\includegraphics[scale=0.6]{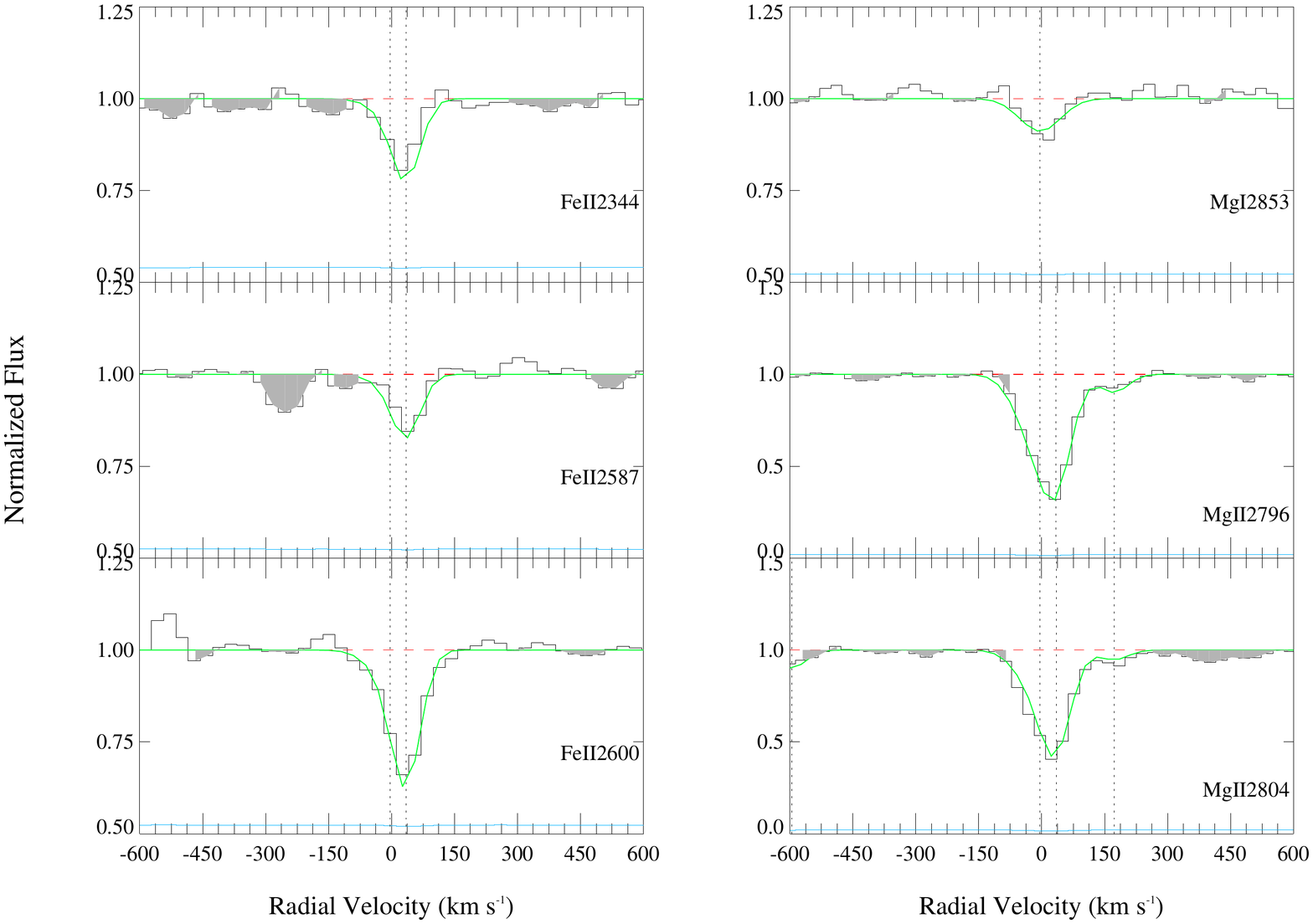}
\caption{ 
Same as Fig. 7, but for metal lines in the $z = 0.4799$ system in the MagE spectrum of Q1355-2257A.
}
\label{fig_1355a_pro}
\end{figure}
\clearpage

\begin{figure}[!th]
\includegraphics[scale=0.6]{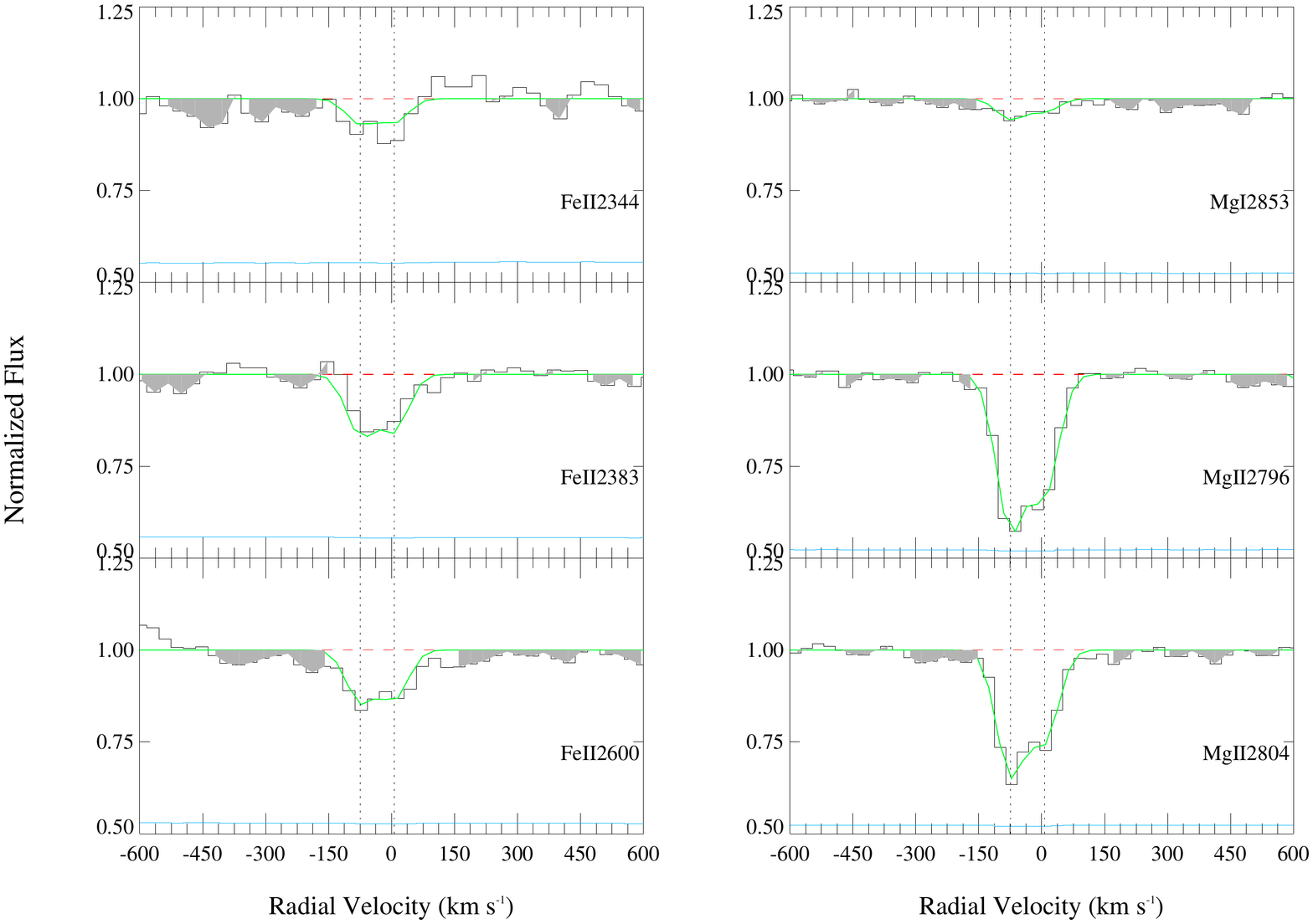}
\caption{ 
Same as Fig. 7, but for metal lines in the $z = 0.4797$ system in the MagE spectrum of Q1355-2257B.
}
\label{fig_1355b_pro}
\end{figure}

\clearpage

\begin{figure}[!th]
\includegraphics[scale=0.6]{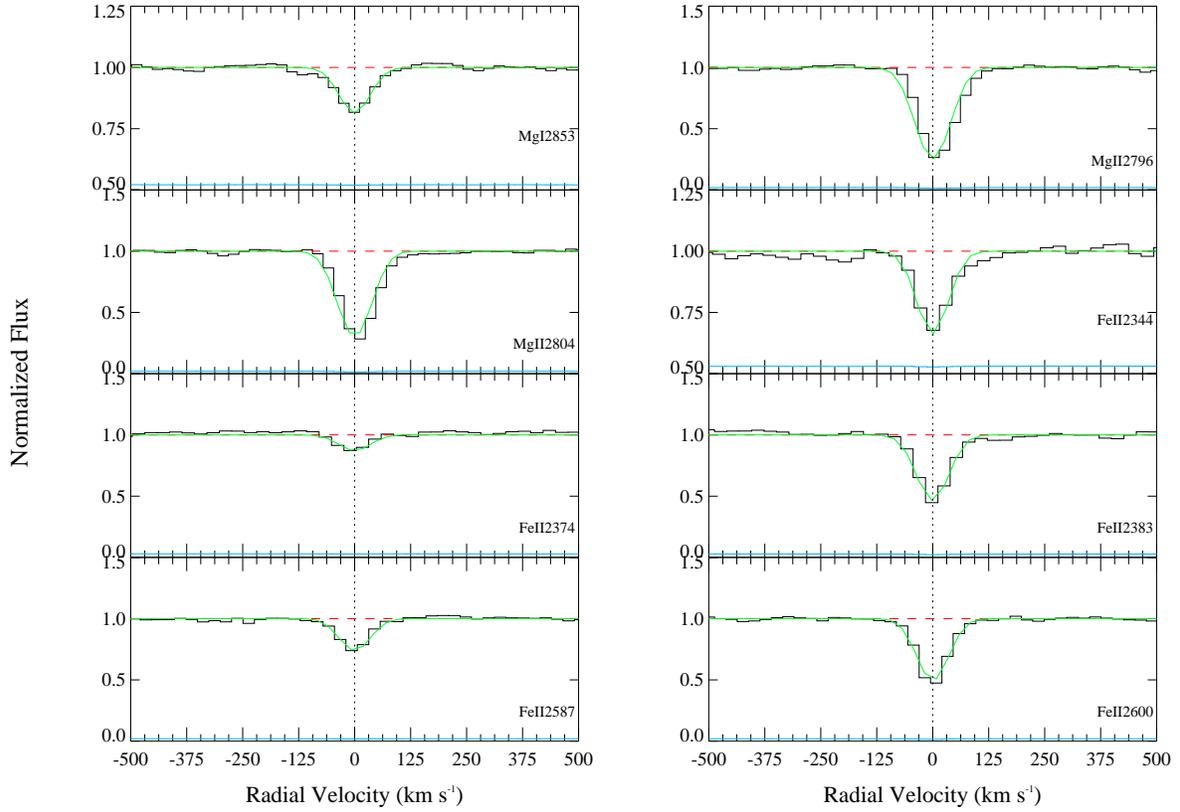}
\caption{ 
Same as Fig. 7, but for metal lines in the $z = 0.7022$ system in the MagE spectrum of Q1355-2257B. No metal lines are detected toward Q1355-2257A.
}
\label{fig_1355b_pro2}
\end{figure}

\clearpage

\begin{figure}[!th]
\includegraphics[scale=0.67, angle=0]{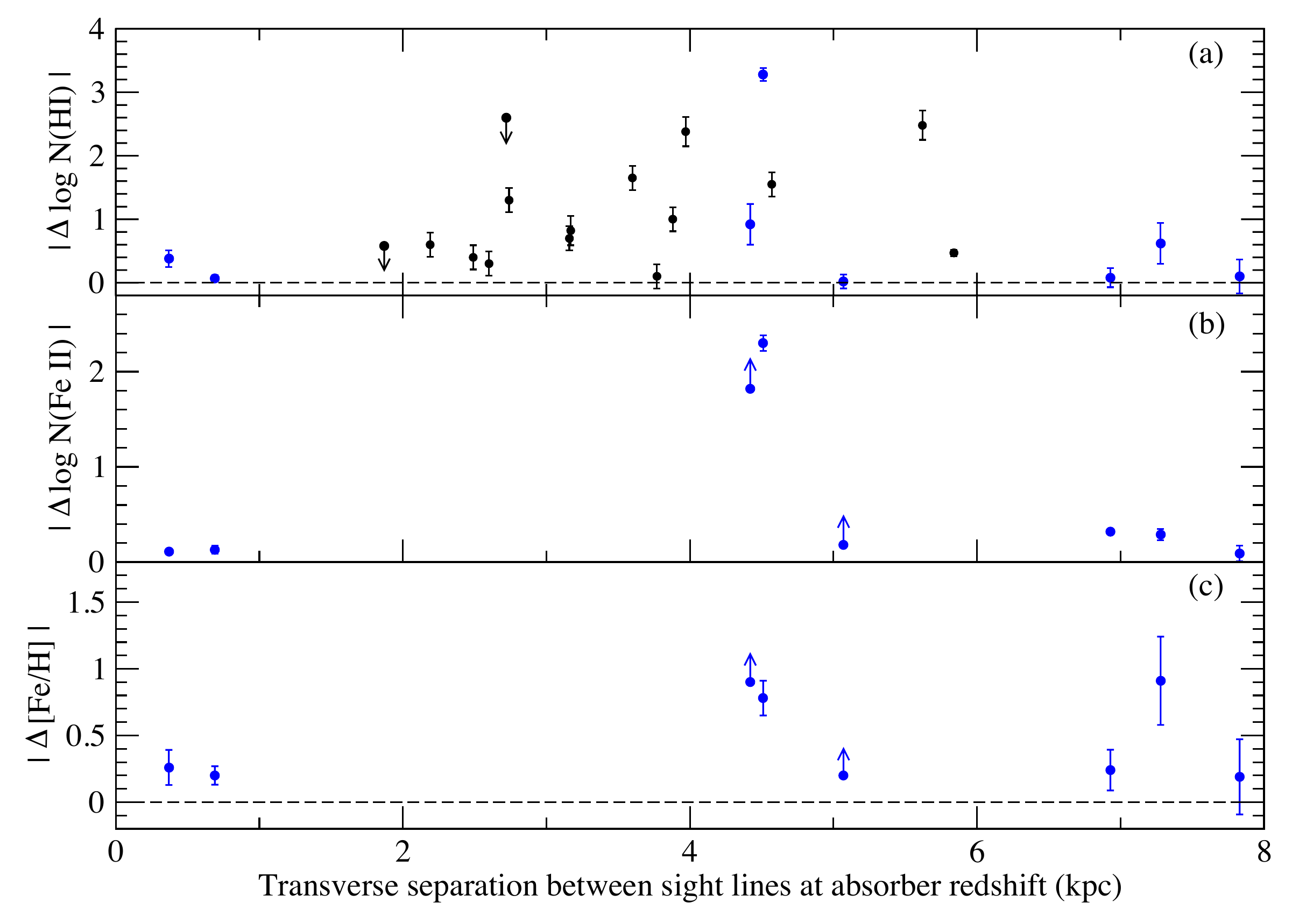}
\caption{ Absolute values of the differences (in dex) in (a) H I column densities, (b) Fe II column densities, and (c) Fe abundances between GLQ sight lines vs. the transverse separation between the sight lines at the absorber redshift. Blue symbols denote absorbers where measurements of both metal and H~I column densities are available, while black symbols denote absorbers with measurements of only H I column densities.}
\label{absdifvssep1}
\end{figure}
\clearpage

\begin{figure}[!th]
\includegraphics[scale=0.67, angle=0]{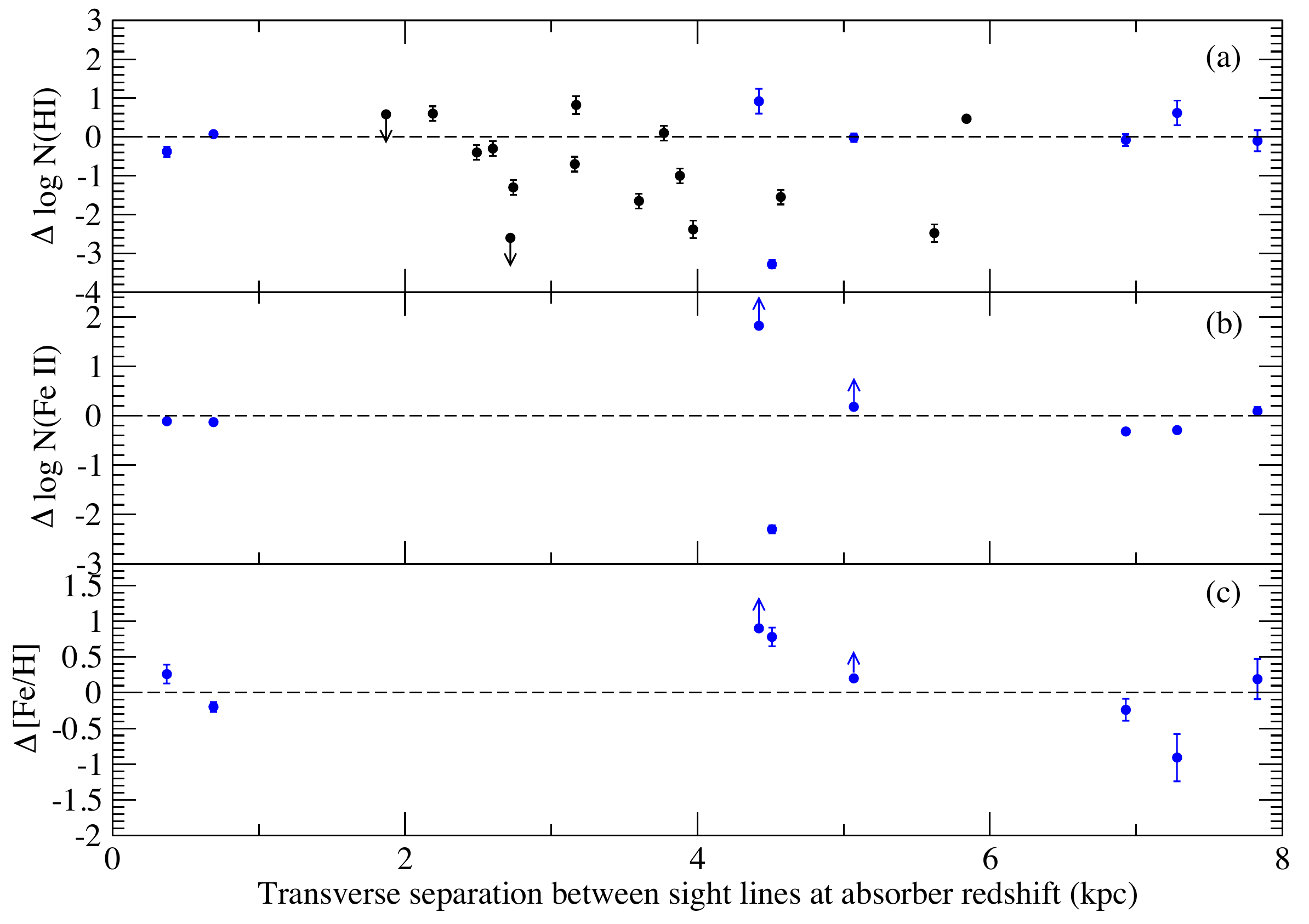}
\caption{ Differences (in dex) in (a) H I column densities, (b) Fe II column densities, and (c) Fe abundances between GLQ sight lines vs. the transverse separation between the sight lines at the absorber redshift. All differences are calculated by subtracting the quantities for the sight line toward the brighter image from those for the sight line toward the fainter image. Blue symbols denote absorbers where measurements of both metal and H I column densities are available, while black symbols denote absorbers with measurements of only H~I column densities.}
\label{absdifvssep_nosign_1}
\end{figure}

\clearpage

\begin{figure}[!th]
\includegraphics[scale=0.67]{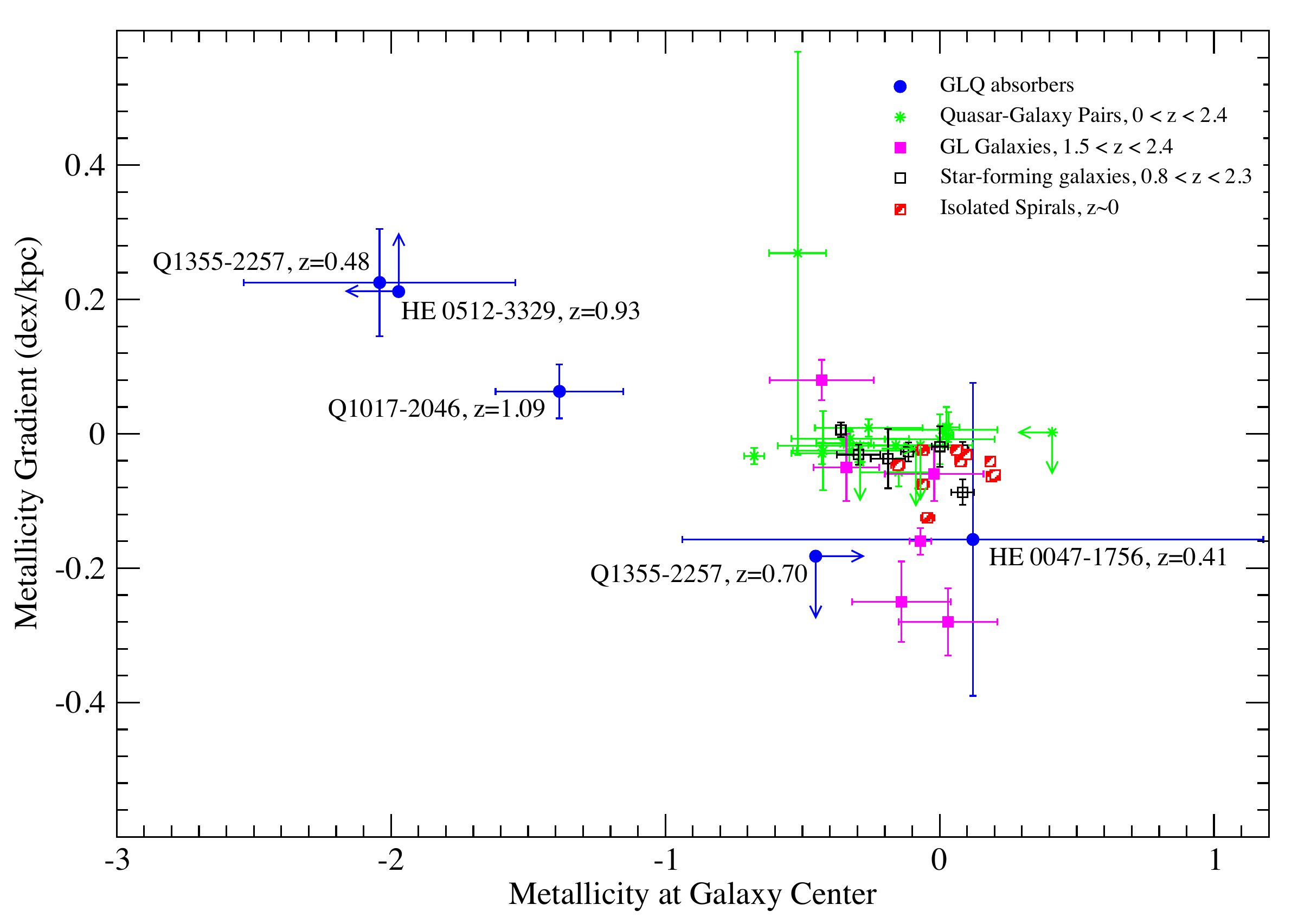}
\caption{ Metal abundance gradient in dex/kpc vs. abundance at the galaxy center for the GLQ absorbers with metal information, compared with measurements  for quasar-galaxy pairs studied with integral field spectroscopy or long-slit spectroscopy, and with other measurements from the literature. The blue circles show the GLQ results for [Fe/H] in absorbers at lens redshifts. The green stars show [O/H] measurements from integral field spectroscopy or long-slit spectroscopy of quasar-galaxy pairs. The magenta squares show [O/H] values for arcs in gravitationally lensed galaxies. The black unfilled squares show [O/H] measurements for star-forming galaxies at $0.8 < z < 2.3$, and the red hatched squares show measurements for isolated spiral galaxies at $z \sim 0$. (See text for further details.)
}
\label{gradvsmeanmet}
\end{figure}

\clearpage

\begin{figure}[!th]
\includegraphics[scale=0.63]{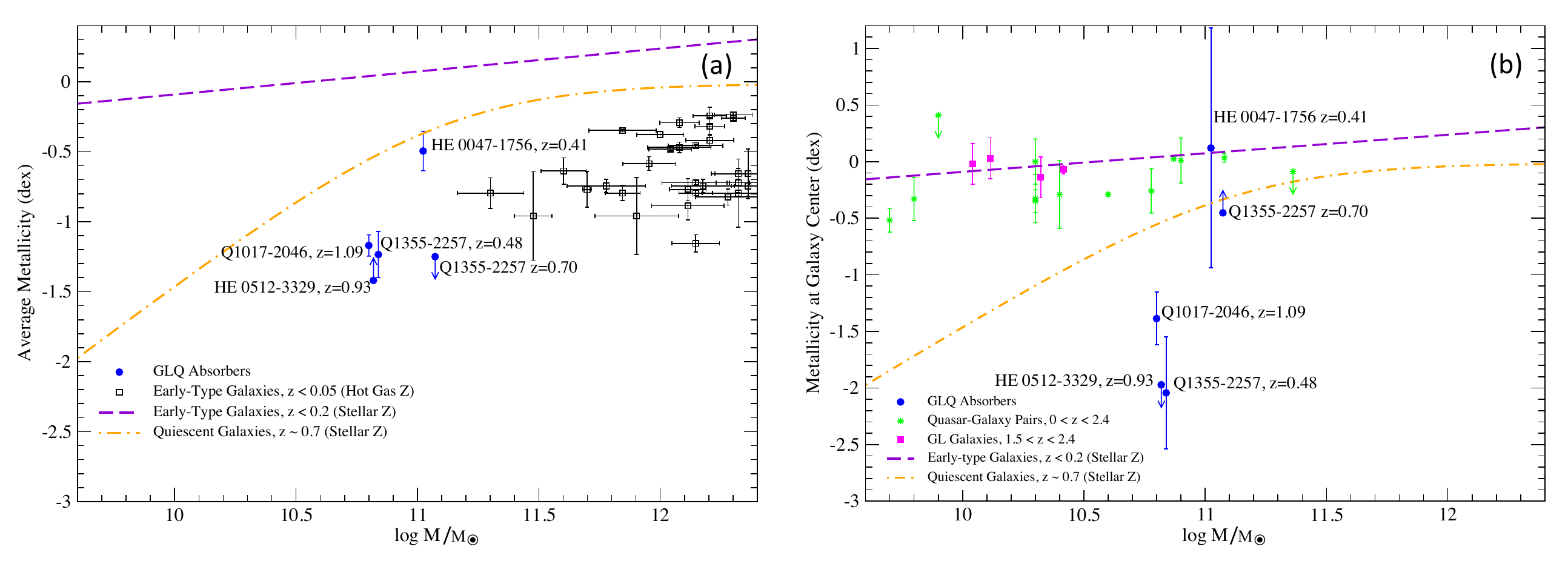}
\caption{Mass-metallicity relation for GLQ absorbers compared with other measurements and/or expected trends. In each panel, the blue circles show the measurements for GLQ absorbers at lens redshifts. GLQ data points in panel (a) show the mean Fe abundance of the two sight lines, while GLQ data points in panel (b) show the Fe abundance inferred to be at the galaxy center.  For Q1355-2257, both possibilities for the lens galaxy redshift are shown. { Black squares in panel (a) show the measurements for early-type galaxies at $z$$<$0.05 based on X-ray observations of hot gas.}  { Green stars and magenta squares in panel (b)} show [O/H] measurements from emission-line spectroscopy for quasar-galaxy pairs at 0$<$$z$$<$2.4, and for arcs in gravitationally lensed galaxies at 1.5$<$$z$$<$2.4, respectively (see text for further details). For reference, { the total mass vs. stellar metallicity relations are shown in each panel for early-type galaxies at $z<0.2$ (violet dashed curve), and for quiescent galaxies at $z \approx 0.7$ (orange dotted-dashed curve).} }
\label{MZR_lens}
\end{figure}

\clearpage

\renewcommand{\baselinestretch}{1.15}
\begin{deluxetable}{llccccc}
\tabletypesize{\footnotesize}
\tablecolumns{8} 
\tablewidth{0pt}
\tablecaption{Targets Observed
}
\tablehead{ 
\colhead{QSO} & 
\colhead{$N_{im}^{a}$} &
\colhead{$z_{QSO}$} &
\colhead{$z_{lens}$} &
\colhead{Mag$_{A}$, Mag$_{B}$} &
\colhead{$ \Delta \theta_{AB}^{b}$}($\arcsec$)&
\colhead{$l_{AB, l}$}(kpc)$^{c}$
}
\startdata
Q1017-2046 & 2&2.545&1.086 &17.43,  19.58 (F555W) 
&0.85&6.9 \\
Q1054+2733 & 2&1.452&0.230 &17.21, 19.22 (V) 
&1.27&4.7\\
Q1349+1227 & 2&1.722&0.645 &17.79, 19.30  (V)
&3.0&20.7\\
Q1355-2257 & 2&1.373&0.48 or 0.70 &17.61, 19.27 (F555W)
&1.22&7.3 or 8.7$^{d}$\\
\enddata
\tablecomments{$a$: Number of lensed quasar images; 
$b$:  Angular separation between lensed quasar images; 
$c$: Transverse separation  between the GLQ sight lines at the lens redshift; 
$d$: The separation 
between the sight lines in the lensing galaxy would be 7.3 kpc if the lens galaxy is at $z=0.48$, or 8.7 kpc if the lens galaxy is at $z=0.70$. (See the text for more details.)}
\label{tab_targsum}
\end{deluxetable}

\renewcommand{\baselinestretch}{1.15}
\begin{deluxetable}{llccccc}
\tabletypesize{\footnotesize}
\tablecolumns{8} 
\tablewidth{0pt}
\tablecaption{Absorbers along the Sight Lines Observed
}
\tablehead{ 
\colhead{QSO} & 
\colhead{$z_{abs}$} &
\colhead{W$^{\rm 2796}_{A}$}({\AA})$^{a}$&
\colhead{W$^{\rm 2796}_{B}$}({\AA})$^{b}$& 
\colhead{$l_{AB, a}$}(kpc)$^{c}$&

\colhead{$r_{A}$(kpc)$^{d}$} &
\colhead{$r_{B}$(kpc)$^{e}$}

}
\startdata
Q1017-2046 &
 1.086&$2.23 \pm 0.01$ &$1.29 \pm 0.02$ &
 6.9&5.4 &1.5\\
Q1054+2733 & 
0.6794&$0.77 \pm 0.10$&$1.07 \pm 0.19$&
1.9&...&...\\
Q1349+1227 & 
1.2366&$2.84 \pm  0.12$& & 
5.8&...&...\\
Q1355-2257$^{f}$ & 
0.4798&$0.75 \pm 0.02$ &$0.55 \pm 0.02$&
7.3  &5.7 &1.6\\
Q1355-2257$^{g}$ & 
0.7022&$< 0.06$& $0.62 \pm 0.02$ &
4.4 or 8.7&6.9 &1.9\\

\enddata
\tablecomments{
$a$: Rest equivalent width of the Mg~II $\lambda 2796$ absorption line in sight line A;  
$b$: Rest equivalent width of the Mg~II $\lambda 2796$ absorption line in sight line B;  
$c$: Transverse separation between the GLQ sight lines at the absorber redshift; 
$d$: Impact parameter of image A from the galaxy, i.e. projected distance between the A sight line and the galaxy center (listed when the absorber is the lensing galaxy itself, i.e. 
$z_{abs} = z_{lens}$).
$e$: Impact parameter of image B from the galaxy, i.e. projected distance between the B sight line and the galaxy center (listed when the absorber is the lensing galaxy itself, i.e. 
$z_{abs} = z_{lens}$).
$f$: The separation 
between the sight lines in the absorbing galaxy at $z_{abs}=0.48$ would be 7.3 kpc {whether the lens galaxy is at $z=0.48$ or at $z=0.70$.} (See the text for more details.)
$g$: The separation 
between the sight lines in the absorbing galaxy at $z_{abs}=0.70$ would be 4.4 kpc if the lens galaxy is at $z=0.48$, or 8.7 kpc if the lens galaxy is at $z=0.70$. (See the text for more details.)
}
\label{tab_targsum}
\end{deluxetable}

\renewcommand{\baselinestretch}{1.15}
\begin{deluxetable}{ccc|cc}
\tabletypesize{\footnotesize}
\tablecolumns{5} 
\tablewidth{0pt}
\tablecaption{Summary of Observations
}
\tablehead{ 
\colhead{QSO} &
\colhead{UV setting} &

\colhead{$t_{exp}$(s)$\times$$n_{exp}^{a}$} &
\colhead{Optical setting} &
\colhead{$t_{exp}$(s)$\times$$n_{exp}^{b}$} 
}
\startdata
Q1017-2046&STIS G230L 2376&1104$\times$4 + 1334$\times$10&Magellan MagE& 2700$\times$2 + 3600$\times$2 (A), \\
&&(A,B)&&3600$\times$2 (B)\\
Q1054+2733&STIS G230L 2376&1143$\times$2 (A,B) &SDSS& 2700$\times$3 (A,B unresolved)\\
Q1349+1227&STIS G230L 2376&1133$\times$1+1134$\times$1 (A,B) &SDSS& 2700$\times$3 (A,B unresolved)\\
Q1355-2257&STIS G230L 2376&1132$\times$1+1133$\times$1+&Magellan MagE&2700$\times$5(A),\\
&&1368$\times$2+1369$\times$2 (A,B)&&2700$\times$1 + 3600$\times$4(B)\\
\enddata
\tablecomments{ $a$: Exposure time and number of individual exposures for HST STIS. $b$: Exposure time and number of individual exposures for the optical spectra obtained with Magellan MagE or adopted from the SDSS.}
\label{tab_obssum}
\end{deluxetable}

\renewcommand{\baselinestretch}{1.15}
\begin{deluxetable}{cclcl}
\tabletypesize{\footnotesize}
\tablecolumns{5} 
\tablewidth{0pt}
\tablecaption{Rest-frame H I Ly-$\alpha$, Ly-$\beta$ equivalent widths for A and B sight lines  
}
\tablehead{ 
\colhead{{ GLQ}} &
\colhead{{$z_{\rm H I}^{A}$}} &

\colhead{{ $W_{Ly\alpha}^{A}$, $W_{Ly\beta}^{A}$({\AA})}} &

\colhead{{ $z_{\rm H I}^{B}$}} &
 \colhead{{ $W_{Ly\alpha}^{B}$, $W_{Ly\beta}^{B}$({\AA})}}
}
\startdata

{ Q1017-2046}&{ 1.08803$\pm$0.00044}&{ 4.22$^{b}$$\pm$0.79, 1.92$^{b}$$\pm$0.60}&{ 1.08714$\pm$0.00053}&{ 3.72$^{b}$$\pm$0.77, 0.82$^{b}$$\pm$0.91}\\
{ Q1054+2733}&{ 0.68153$\pm$0.00036}&{ 1.47$\pm$0.49}, N/A&{ 0.68350$\pm$0.00286}&{ $<$ 2.47$^{a}$}, N/A\\
{ Q1349+1227}&{ 1.23750$\pm$0.00026}&{ 17.71$\pm$2.65, 2.36$\pm$0.61}&{ 1.24022$\pm$0.00083}&{ 28.05$\pm$4.25, 2.21$\pm$0.82}\\
{ Q1355-2257}&{ 0.48169$\pm$0.00048}&{ 1.70$\pm$0.51}, N/A&{ 0.48128$\pm$0.00112}&{ 2.98$\pm$1.42}, N/A\\
{ Q1355-2257}&{ 0.70721$\pm$0.00045}&{ 0.86$^{b}$$\pm$0.28}, N/A&{ 0.70501$\pm$0.00067}&{ 2.19$\pm$0.84}, N/A\\

\enddata
\tablecomments{
{ $a$: 3 $\sigma$ upper limit
$b$: Equivalent width measurement affected by blends and/or noise.}}
\label{tab_HI}
\end{deluxetable}

\renewcommand{\baselinestretch}{1.15}
\begin{deluxetable}{ccccc}
\tabletypesize{\footnotesize}
\tablecolumns{7} 
\tablewidth{0pt}
\tablecaption{H I column density measurements 
{ based on Voigt Profile Fitting}
}
\tablehead{ 
\colhead{GLQ} &
\colhead{$z_{\rm H I}^{A}$} &
\colhead{log $N_{\rm H I}^{A}$} &
\colhead{$z_{\rm H I}^{B}$} &
\colhead{log $N_{\rm H I}^{B}$ } 
}
\startdata
Q1017-2046A,B&1.08803$\pm$0.00044 
& 19.87$\pm$0.09&1.08714$\pm$0.00053 
& 19.79$\pm$0.12\\
Q1054+2733A,B&0.68153$\pm$0.00036 
&{ 18.48$\pm$0.18}&0.68350$\pm$0.00286&{ $<$ 19.06}\\
Q1349+1227A,B&1.23750$\pm$0.00026  
&20.90$\pm$0.02&1.24022$\pm$0.00083 
&21.37$\pm$0.05\\
Q1355-2257A,B&{ 0.48183$\pm$0.00039}
& {18.81$\pm$0.18}&{0.48153$\pm$0.00116} 
& 19.43$\pm$0.27\\
Q1355-2257A,B& 0.70725$\pm$ 0.00064  
&18.20$\pm$0.23& 0.70534$\pm$ 0.00079 
& 19.12$\pm$0.23\\
\enddata
\tablecomments{ The  H I column densities $N_{\rm H I}$ are in units of cm$^{-2}$.
}
\label{tab_HI}
\end{deluxetable}

\begin{landscape}		
\renewcommand{\baselinestretch}{1.15}
\begin{deluxetable}{ccccccccc}
\tabletypesize{\footnotesize}
\tablecolumns{5} 
\tablewidth{0pt}
\tablecaption{
 Results of Voigt Profile Fitting for Ions in the $z = 1.0859$ Absorber in the Sight Line toward Q1017-2046A
}

\tablehead{ 
\colhead{$z$} &
\colhead{$b_{eff}$} &
\colhead{log $N_{\rm Mg I}$} &
\colhead{log $N_{\rm Mg II}$} &
\colhead{log $N_{\rm Al II}$} &
\colhead{log $N_{\rm Al III}$} &
\colhead{log $N_{\rm Si II}$}& 
\colhead{log $N_{\rm Fe II}$}&
\colhead{log $N_{\rm C IV}$}
}

\startdata
1.085905 & 55.32  $\pm$ 0.92& 12.39$\pm$0.04 & 14.01$\pm$0.01 & 13.12$\pm$0.07  &
 12.64$\pm$0.32 & 14.45$\pm$0.16 & 14.03$\pm$0.01 & ...\\
$\pm$0.000005 &&&&&&&&\\
1.086955& 16.30 $\pm$ 0.72 & 11.91$\pm$0.07 & 13.83$\pm$0.04 & 13.21$\pm$0.28  &
13.33$\pm$0.10 & 14.87$\pm$0.17 & 13.94$\pm$0.03 &...\\
$\pm$0.000005&&&&&&&&\\
1.088699 & 10.65  $\pm$ 1.69& 11.90$\pm$0.10 & 13.26$\pm$0.06 & 12.44$\pm$0.44  &
12.27$\pm$0.55 & 15.26$\pm$0.09 & 13.20$\pm$0.05 & ...\\
$\pm$0.000009 &&&&&&&&\\
1.089270 & $10.00$        &       ...      & 12.40$\pm$0.09 &       ...       &      ...      &       ...      &       ...    &...  \\
$\pm$0.000049&&&&&&&&\\
\hline
 1.086822 & $36.00$ &..&..&..&..&...&...&14.41$\pm$0.05\\
$\pm$0.000016&&&&&&&&\\
1.088587 & 42.69  $\pm$ 7.91 &...&...&...&...&...&...&13.91$\pm$0.04 \\
$\pm$0.000028&&&&&&&&\\
1.089569 & 60.58  $\pm$ 19.68&...&...&...&...&...&...&13.69$\pm$0.15 \\
$\pm$0.000121&&&&&&&&\\

\enddata
\tablecomments{The effective Doppler parameters $b_{eff}$  are in units of km s$^{-1}$ and the column densities $N$ are in units of cm$^{-2}$.}
\label{tab_vp1017a}
\end{deluxetable}

\renewcommand{\baselinestretch}{1.15}
\begin{deluxetable}{cccccccc}
\tabletypesize{\footnotesize}
\tablecolumns{6} 
\tablewidth{0pt}
\tablecaption{
Results of Voigt Profile Fitting for Ions in the $z = 1.086$ Absorber   in the Sight Line toward Q1017-2046B
}

\tablehead{ 
\colhead{$z$} &
\colhead{$b_{eff}$} &
\colhead{log $N_{\rm Mg I}$} &
\colhead{log $N_{\rm Mg II}$} &
\colhead{log $N_{\rm Si II}$} &
\colhead{log $N_{\rm Fe II}$} &
\colhead{log $N_{\rm Al II}$}&
\colhead{log $N_{\rm Al III}$}
}
\startdata
1.085979$\pm$0.000006 & 18.07  $\pm$ 0.78 & 12.21$\pm$0.06 & 13.74$\pm$0.03 & 14.30$\pm$0.39 & 13.91$\pm$0.02 & 12.86$\pm$0.10&...\\
1.086504$\pm$0.000029 & 41.56  $\pm$ 2.91 & 11.68$\pm$0.20 & 13.22$\pm$0.03 & 14.54$\pm$0.17 & 12.99$\pm$0.05 &12.32$\pm$0.19&...\\
1.087443$\pm$0.000015 & 8.00        & 11.63$\pm$0.21 & 13.08$\pm$0.07 &       ...      & 12.91$\pm$0.07&11.65$\pm$0.78&... \\
1.088767$\pm$0.000058 & 11.00        &       ...      & 12.24$\pm$0.10 &       ...      &       ...    & ... &... \\
\hline
 1.086534$\pm$0.000091 & 62.79 $\pm$ 22.94&...&...&...&...&...& 12.93$\pm$0.12 \\

\enddata
\tablecomments{The effective Doppler parameters $b_{eff}$  are in units of km s$^{-1}$ and the column densities $N$ are in units of cm$^{-2}$.}
\label{tab_vp1017b}
\end{deluxetable}
\end{landscape}
\clearpage

\renewcommand{\baselinestretch}{1.15}
\begin{deluxetable}{ccc|cc}
\tabletypesize{\footnotesize}
\tablecolumns{5} 
\tablewidth{0pt}
\tablecaption{
Total Column Densities for the $z = 1.086$ Absorbers  in the Sight Lines toward Q1017-2046A and Q1017-2046B
}
\tablehead{ 
\colhead{Ion} &
\colhead{log $N^{\rm fit}_{A}$(cm$^{-2}$)} &
\colhead{log $N^{\rm AOD}_{A}$(cm$^{-2}$)} &
\colhead{log $N^{\rm fit}_{B}$(cm$^{-2}$)} &
\colhead{log $N^{\rm AOD}_{B}$(cm$^{-2}$)} 
}
\startdata
Mg {\sc i}  & $12.61\pm0.03$ & $12.61\pm0.02$ & $12.40\pm0.06$ & $12.35\pm0.03$\\
Mg {\sc ii} & $14.28\pm0.02$ & $14.23\pm0.01$ &$13.93\pm0.02$ & $13.83\pm0.01$\\
Al {\sc ii} & $13.51\pm0.15$ &       ...      & $12.99\pm0.09$ &       ...     \\
Al {\sc iii}& $13.44\pm0.10$ &       ...      & $12.93\pm0.12$ &       ...    \\
Si {\sc ii} & $15.46\pm0.08$ &       ...     & $14.74\pm0.18$ &       ...      \\
Fe {\sc ii} & $14.32\pm0.02$ & { 14.28$\pm0.01$} & $14.00\pm0.02$ & $13.97\pm0.07$\\
C {\sc iv}  & $14.59\pm0.04$ &       ...     &...&... \\
\enddata
\label{tab_total1017a}
\end{deluxetable}

\renewcommand{\baselinestretch}{1.15}
\begin{deluxetable}{ccc}
\tabletypesize{\footnotesize}
\tablecolumns{3} 
\tablewidth{0pt}
\tablecaption{
Element Abundances Relative to Solar for the $z = 1.086$ Absorbers  in the Sight Lines toward Q1017-2046A and Q1017-2046B
}
\tablehead{ 
\colhead{Element} &
\colhead{[X/H]$_{A}$} &
\colhead{[X/H]$_{B}$} 
}
\startdata
Mg &  -1.18$\pm$0.09& -1.45$\pm$0.12\\
Al &  -0.54$\pm$0.13& -0.98$\pm$ 0.14\\
Si &  0.08$\pm$0.12 &-0.56$\pm$0.21\\
Fe &  -1.05$\pm$0.09& -1.29$\pm$0.12\\
\enddata
\\
\label{tab_1017ab_x/h}
\end{deluxetable}
	
\renewcommand{\baselinestretch}{1.15}
\begin{deluxetable}{cccc}
\tabletypesize{\footnotesize}
\tablecolumns{5} 
\tablewidth{0pt}
\tablecaption{
 { Results of Voigt Profile Fitting for Ions in the $z = 1.0859$ Absorber toward Q1017-2046A+B from archival spatially unresolved Keck HIRES data}
}

\tablehead{ 
\colhead{ $z$} &
\colhead{ $b_{eff}$} &
\colhead{{ log $N_{\rm Mg II}$} }&
\colhead{{ log $N_{\rm Fe II}$}}
}

\startdata
{ 1.085403$\pm$0.000007} &  4.60  $\pm$ 1.30 & { 13.66$\pm$0.06  }& { 13.04$\pm$0.04 }\\
{ 1.085534$\pm$0.000017 }&   6.98  $\pm$ 4.38 & { 12.77$\pm$0.02 }& { 12.59$\pm$0.06} \\
{ 1.085876$\pm$0.000004 }&  15.47  $\pm$ 0.68 & { 13.94$\pm$0.03 }& { 13.85$\pm$0.01} \\
{ 1.086063$\pm$0.000006} &   4.75  $\pm$  1.03& { 13.35$\pm$0.08 }& { 13.28$\pm$0.04} \\
{ 1.086176$\pm$0.000010 }&   3.89 $\pm$  2.82& { 13.32$\pm$0.07 }&{ 12.70$\pm$0.06} \\
{ 1.086317$\pm$0.000006 }&   4.28  $\pm$  1.33& { 13.24$\pm$0.06 }& { 12.99$\pm$0.06 }\\
{ 1.086888$\pm$0.000067} &  32.74  $\pm$ 4.02& { 12.88$\pm$0.04 }&  { ...         }   \\
{ 1.086489$\pm$0.000005 }&  13.78 $\pm$  1.12& { 13.57$\pm$0.02 }&  { 13.30$\pm$0.02} \\
{ 1.086997$\pm$0.000008 }&   6.68  $\pm$ 0.59& { 13.92$\pm$0.08 }&  { 13.80$\pm$0.03} \\
{ 1.087192$\pm$0.000008 }&   6.96  $\pm$ 0.57& { 14.82$\pm$0.07 }& { 13.75$\pm$0.02} \\
{ 1.088701$\pm$0.000004} &   3.16  $\pm$  0.64& { 14.04$\pm$0.10 }&  { 12.99$\pm$0.06} \\
{ 1.088927$\pm$0.000001} &  4.90  $\pm$ 0.58& { 13.52$\pm$0.05 }&  { 13.25$\pm$0.04} \\
{ 1.089448$\pm$0.000020} &   3.02 $\pm$ 0.78& { 12.67$\pm$0.05 }&  { 12.16$\pm$0.15} \\
\enddata
\tablecomments{{ The effective Doppler parameters $b_{eff}$  are in units of km s$^{-1}$ and the column densities $N$ are in units of cm$^{-2}$.}}
\label{tab_vp1017abhires}
\end{deluxetable}

\renewcommand{\baselinestretch}{1.15}
\begin{deluxetable}{ccc}
\tabletypesize{\footnotesize}
\tablecolumns{3} 
\tablewidth{0pt}
\tablecaption{
{ Total Column Densities for the $z = 1.0859$ Absorber in the Sight Lines toward Q1017-2046A+B from archival spatially unresolved Keck HIRES data)}
}
\tablehead{ 
\colhead{{ Ion}} &
\colhead{{ log $N^{\rm fit}_{A+B}$(cm$^{-2}$)}} &
\colhead{{ log $N^{\rm AOD}_{A+B}$(cm$^{-2}$)}} 
}
\startdata
{ Mg {\sc ii} }& { 15.05$\pm$0.04} & { 14.37$\pm$0.01} \\
{ Fe {\sc ii} }& { 14.46$\pm$0.01} & { 14.33$\pm$0.01} \\
\enddata
\label{tab_total1017abhires}
\end{deluxetable}

\renewcommand{\baselinestretch}{1.15}
\begin{deluxetable}{cccc}
\tabletypesize{\footnotesize}
\tablecolumns{5} 
\tablewidth{0pt}
\tablecaption{Results of Voigt Profile Fitting for Ions in the $z = 0.6794$ Unresolved Absorber in the Sight Line toward Q1054+2733
}
\tablehead{ 
\colhead{$z$} &
\colhead{$b_{eff}$(km s$^{-1}$)} &
\colhead{log $N_{\rm Mg II}$(cm$^{-2}$)} &
\colhead{log $N_{\rm Fe II}$(cm$^{-2}$)} 
}
\startdata
{ 0.679410$\pm$0.000029} & 68.59 $\pm$ 14.97 & { 13.29$\pm$0.03} & { 13.49$\pm$0.04} \\
\enddata
\label{tab_vp1054ab}
\end{deluxetable}

\renewcommand{\baselinestretch}{1.15}
\begin{deluxetable}{ccc}
\tabletypesize{\footnotesize}
\tablecolumns{3} 
\tablewidth{0pt}
\tablecaption{
 Total Column Densities for the $z = 0.6794$ Unresolved Absorber in the Sight Line toward Q1054+2733
}
\tablehead{ 
\colhead{Ion} &
\colhead{log $N^{\rm fit}$(cm$^{-2}$)} &
\colhead{log $N^{\rm AOD}$(cm$^{-2}$)} 
}
\startdata
Mg {\sc ii} & { 13.29$\pm$0.03} & { 13.44$\pm$0.04} \\
Fe {\sc ii} & { 13.49$\pm$0.04}& { 13.50$\pm$0.04} \\
\enddata
\label{tab_total1054ab}
\end{deluxetable}

\renewcommand{\baselinestretch}{1.15}
\begin{deluxetable}{cc}
\tabletypesize{\footnotesize}
\tablecolumns{2} 
\tablewidth{0pt}
\tablecaption{
Element Abundances Relative to Solar for the $z = 0.6794$ Unresolved Absorber in the Sight Line toward Q1054+2733
}
\tablehead{ 
\colhead{Element} &
\colhead{[X/H]} 
}
\startdata
Mg & { $>${ -1.02}} \\
Fe & { $>${ -0.86}} \\
\enddata
\\
\tablecomments{Approximate mean abundance limits corresponding to the upper limit on the mean H I column density.}
\label{tab_1054ab_x/h}
\end{deluxetable}

\renewcommand{\baselinestretch}{1.15}
\begin{deluxetable}{lccccccc}
\tabletypesize{\footnotesize}
\tablecolumns{7} 
\tablewidth{0pt}
\tablecaption{Results of Voigt Profile Fitting for Ions in the $z = 1.2366$ Unresolved Absorber in the Sight Line toward Q1349+1227
}
\tablehead{ 
\colhead{$z$} &
\colhead{$b_{eff}$}&
\colhead{log $N_{\rm Mg I}$} &
\colhead{log $N_{\rm Mg II}$} &
\colhead{log $N_{\rm Al II}$} &
\colhead{log $N_{\rm Al III}$} &
\colhead{log $N_{\rm Ca II}$} &
\colhead{log $N_{\rm Fe II}$} \\
&(km s$^{-1}$)&(cm$^{-2}$)&(cm$^{-2}$)&(cm$^{-2}$)&(cm$^{-2}$)&(cm$^{-2}$)
}
\startdata
{ 1.234744} &{  23.62} & { 12.27} & { 14.55}& 
{ 13.04} &
{ 12.59} & { 12.59} & { 14.45} \\
{ $\pm$0.000022} & $\pm$ 1.72&{$\pm$0.12}&{$\pm$0.11}&{$\pm$0.19}&{$\pm$0.17}&{$\pm$0.13}&{$\pm$0.08}\\
&&&&&&&\\
{ 1.238496} & { 23.71}& { 12.72}& { 14.88}& 
{ 14.23}&
{ 13.18} & { 12.64} & { 14.55} \\
{ $\pm$0.000021} & $\pm$ 1.62&{$\pm$0.08 }&{$\pm$0.13 }&{$\pm$0.34}&{$\pm$0.07}&{$\pm$0.09}&{$\pm$0.09}\\

\enddata
\label{tab_vp1349a}
\end{deluxetable}

\renewcommand{\baselinestretch}{1.15}
\begin{deluxetable}{ccc}
\tabletypesize{\footnotesize}
\tablecolumns{3} 
\tablewidth{0pt}
\tablecaption{
 Total Column Densities for the $z = 1.2366$ Unresolved Absorber in the Sight Line toward Q1349+1227
}
\tablehead{ 
\colhead{Ion} &
\colhead{log $N^{\rm fit}$(cm$^{-2}$)} &
\colhead{log $N^{\rm AOD}$(cm$^{-2}$)} 
}
\startdata
Mg {\sc i}  & { 12.85$\pm$0.07} & $12.59\pm0.06$ \\
Mg {\sc ii} & { 15.05$\pm$0.09} &       ...      \\
Al {\sc ii} & { 14.25$\pm$0.32} &  ...       \\
Al {\sc iii}& $13.28\pm0.06$ &   ...   \\
{ Ca {\sc ii}} & { 12.91$\pm$0.08} & { 12.80 $\pm$ 0.09}      \\
Fe {\sc ii} & $14.80\pm0.06$ & $14.59\pm0.05$ \\
\enddata
\label{tab_total1349a}
\end{deluxetable}

\renewcommand{\baselinestretch}{1.15}
\begin{deluxetable}{cc}
\tabletypesize{\footnotesize}
\tablecolumns{2} 
\tablewidth{0pt}
\tablecaption{
Element Abundances Relative to Solar for the $z = 1.2366$ Unresolved Absorber  in the Sight Line toward Q1349+1227
}
\tablehead{ 
\colhead{Element} &
\colhead{[X/H]} 
}
\startdata
Mg & { -1.75$\pm$0.10}\\
Al & { -1.35$\pm$ 0.29} \\
{ Ca} & { -2.74$\pm$0.10} \\
Fe & { -2.11}$\pm${ 0.06} \\
\enddata
\tablecomments{Approximate mean abundances corresponding to the mean H I column density.}
\\
\label{tab_1349a_x/h}
\end{deluxetable}	

\renewcommand{\baselinestretch}{1.15}
\begin{deluxetable}{ccccc}
\tabletypesize{\footnotesize}
\tablecolumns{5} 
\tablewidth{0pt}
\tablecaption{Results of Voigt Profile Fitting for Ions in the $z = 0.4799$ Absorber in the Sight Line toward Q1355-2257A
}
\tablehead{ 
\colhead{$z$} &
\colhead{$b_{eff}$(km s$^{-1}$)} &
\colhead{log $N_{\rm Mg I}$(cm$^{-2}$)} &
\colhead{log $N_{\rm Mg II}$(cm$^{-2}$)} &
\colhead{log $N_{\rm Fe II}$(cm$^{-2}$)} 
}
\startdata
$0.479971\pm0.000006$ &  13.43 $\pm$ 1.52 &       ...      & $14.13\pm0.08$ & $13.45\pm0.05$ \\
$0.479781\pm0.000021$ &  51.33 $\pm$ 3.02 & $11.95\pm0.06$ & $13.20\pm0.02$ & $12.74\pm0.12$ \\
$0.480653\pm0.000029$ & 12.00           &       ...      & $12.31\pm0.06$ &       ...      \\
\enddata
\label{tab_vp1355a}
\end{deluxetable}

\renewcommand{\baselinestretch}{1.15}
\begin{deluxetable}{ccccc}
\tabletypesize{\footnotesize}
\tablecolumns{5} 
\tablewidth{0pt}
\tablecaption{
 Results of Voigt Profile Fitting for Ions in the $z = 0.4797$ Absorber in the Sight Line toward Q1355-2257B
}
\tablehead{ 
\colhead{$z$} &
\colhead{$b_{eff}$(km s$^{-1}$)} &
\colhead{log $N_{\rm Mg I}$(cm$^{-2}$)} &
\colhead{log $N_{\rm Mg II}$(cm$^{-2}$)} &
\colhead{log $N_{\rm Fe II}$(cm$^{-2}$)} 
}
\startdata
$0.479634\pm0.000007$ &  11.99  $\pm$ 1.12 & $11.59\pm0.11$ & $13.44\pm0.03$ & $12.96\pm0.05$ \\
$0.480035\pm0.000009$ & 10.26 $\pm$ 1.32 & $11.38\pm0.17$ & $13.20\pm0.03$ & $12.93\pm0.05$ \\
\enddata
\label{tab_vp1355b}
\end{deluxetable}

\renewcommand{\baselinestretch}{1.15}
\begin{deluxetable}{ccc|cc}
\tabletypesize{\footnotesize}
\tablecolumns{5} 
\tablewidth{0pt}
\tablecaption{
 Total Column Densities for the $z = 0.48$ Absorbers in the Sight Lines toward Q1355-2257A and Q1355-2257B
}
\tablehead{ 
\colhead{Ion} &
\colhead{log $N^{\rm fit}_{A}$(cm$^{-2}$)} &
\colhead{log $N^{\rm AOD}_{A}$(cm$^{-2}$)} &
\colhead{log $N^{\rm fit}_{B}$(cm$^{-2}$)} &
\colhead{log $N^{\rm AOD}_{B}$(cm$^{-2}$)} 
}
\startdata
Mg {\sc i}  & $11.95\pm0.06$ & $11.87\pm0.10$ & $11.80\pm0.10$ & $11.63\pm0.10$\\
Mg {\sc ii} & $14.18\pm0.07$ & $13.60\pm0.02$ & $13.64\pm0.02$ & $13.37\pm0.02$\\
Fe {\sc ii} & $13.53\pm0.05$ & $13.42\pm0.06$ & { 13.24}$\pm0.04$ & $13.19\pm0.04$\\
\enddata
\label{tab_total1355ab}
\end{deluxetable}

\renewcommand{\baselinestretch}{1.15}
\begin{deluxetable}{ccc}
\tabletypesize{\footnotesize}
\tablecolumns{2} 
\tablewidth{0pt}
\tablecaption{
Element Abundances Relative to Solar for the $z = 0.48$ absorbers in the Sight Lines toward Q1355-2257A and Q1355-2257B 
}
\tablehead{ 
\colhead{Element} &
\colhead{[X/H]$_{A}$} &
\colhead{[X/H]$_{B}$} 
}
\startdata
Mg & ${ -0.22} \pm { 0.19}$ & ${ -1.38} \pm { 0.27}$ \\
Fe & ${ -0.78} \pm { 0.19}$ & ${ -1.69} \pm { 0.27}$\\
\enddata
\\
\label{tab_1355ab_x/h}
\end{deluxetable}

\renewcommand{\baselinestretch}{1.15}
\begin{deluxetable}{ccc}
\tabletypesize{\footnotesize}
\tablecolumns{3} 
\tablewidth{0pt}
\tablecaption{
 { 3${\sigma}$ Upper Limits for Ions in the $z = 0.7022$ Absorber in the Sight Line toward Q1355-2257A}
}
\tablehead{ 
\colhead{{log $N_{\rm Mg I}$(cm$^{-2}$)}} &
\colhead{{log $N_{\rm Mg II}$(cm$^{-2}$)}} &
\colhead{{ log $N_{\rm Fe II}$(cm$^{-2}$)}} 
}
\startdata
{ $<${ 11.13}} & { $<$ 11.79} & { $<$12.00} \\
\enddata
\label{tab_1355a}
\end{deluxetable}

\renewcommand{\baselinestretch}{1.15}
\begin{deluxetable}{cc}
\tabletypesize{\footnotesize}
\tablecolumns{2} 
\tablewidth{0pt}
\tablecaption{
{ 3${\sigma}$ Upper Limits for Element Abundances Relative to Solar for the $z = 0.7022$ Absorber in the Sight Line toward Q1355-2257A}
}
\tablehead{ 
\colhead{{ Element}} &
\colhead{{ [X/H]}} 
}
\startdata
{ Mg} &  $<$  -1.92 \\
{ Fe} & $<$ -1.70 \\
\enddata
\\
\label{tab_1355a_x/h}
\end{deluxetable}		
		
\renewcommand{\baselinestretch}{1.15}
\begin{deluxetable}{ccccc}
\tabletypesize{\footnotesize}
\tablecolumns{5} 
\tablewidth{0pt}
\tablecaption{
 Results of Voigt Profile Fitting for Ions in the $z = 0.7022$ Absorber in the Sight Line toward Q1355-2257B
}
\tablehead{ 
\colhead{$z$} &
\colhead{$b_{eff}$(km s$^{-1}$)} &
\colhead{log $N_{\rm Mg I}$(cm$^{-2}$)} &
\colhead{log $N_{\rm Mg II}$(cm$^{-2}$)} &
\colhead{log $N_{\rm Fe II}$(cm$^{-2}$)} 
}
\startdata
$0.702257\pm0.0000046$ &  18.30 $\pm$ 0.94 & $12.15\pm0.04$ & $14.40\pm0.04$ & $13.82\pm0.02$ \\
\enddata
\label{tab_vp1355b}
\end{deluxetable}

\renewcommand{\baselinestretch}{1.15}
\begin{deluxetable}{cc}
\tabletypesize{\footnotesize}
\tablecolumns{2} 
\tablewidth{0pt}
\tablecaption{
Element Abundances Relative to Solar for the $z = 0.7022$ Absorber  in the Sight Line  toward Q1355-2257B
}
\tablehead{ 
\colhead{Element} &
\colhead{[X/H]} 
}
\startdata
Mg & { -0.32}$\pm${ 0.24}\\
Fe & { -0.80}$\pm${ 0.23}\\
\enddata
\\
\label{tab_1355b_x/h}
\end{deluxetable}

\begin{landscape}	
\renewcommand{\baselinestretch}{1.15}
\begin{deluxetable}{lccccccccc}
\tabletypesize{\footnotesize}
\tablecolumns{9} 
\tablewidth{0pt}
\tablecaption{{ Absorbers with H I and/or Fe II Column Density Measurements toward GLQs}
}

\tablehead{ 
\colhead{{ GLQ  Sight Lines} }&
\colhead{{ $z_{qso}$} }&
\colhead{{ $z_{lens}$} }&
\colhead{{ $z_{abs}$} }&
\colhead{{ $l_{A,B, a}^{1}$}}&
\colhead{{ $\Delta$ log $N_{\rm H I}$}} &
\colhead{{ $\Delta$ log $N_{\rm Fe II}$}} &
\colhead{{ $\Delta$[Fe/H]/$\Delta r^{2}$} }&
\colhead{{ [Fe/H]$_{0}^{3}$}}&
\colhead{{ Ref.$^{4}$}}
}
\startdata

{Q1017-2046A,B}&2.545&1.085&1.085&6.93&{ -0.08}$\pm$ 0.15&$-0.32 \pm 0.03$&{ 0.063}$\pm$ 0.040&  -1.386$\pm$0.233&1\\
{Q1355-2257A,B$^{5}$}&1.373&0.48&0.48&7.28&{ 0.62} $\pm$ { 0.32}&{ -0.29}$ \pm$ 0.06&{ 0.220}$ \pm$  { 0.080}&{ -2.042}$\pm${ 0.495}&1\\
{Q1355-2257A,B$^{5}$}&1.373&0.70&0.70&8.73& 0.92$ \pm$ 0.32&$>1.82$& $<${ -0.182}&$>${ -0.452}&1\\
{Q0047-1756A,B}&1.676&0.408&0.408&7.83&$-0.1 \pm 0.27$&$0.09 \pm 0.082$&$-0.157 \pm 0.233$ & $0.121 \pm 1.059$ &2\\
{Q0512-3329A,B}&1.587&0.933&0.933&5.07&$-0.02 \pm 0.11$&$>0.18$&$>0.217$&$<-1.972$&3\\
&&&&&&&&&\\
{Q1054+2733A,B}&1.452&0.23&0.6794&1.87&$<${ 0.58}&NA&NA&NA&1\\
{Q1349+1227A,B}&1.722&0.645&1.2366&5.84&$0.47 \pm 0.05$&NA&NA&NA&1\\
{H1413+117A,B}&2.551&1.0&1.44&3.17&$0.82 \pm 0.23$&NA&NA&NA&4\\
{H1413+117A,C}&2.551&1.0&1.44&3.60&$-1.65 \pm 0.19$&NA&NA&NA&4\\
{H1413+117A,D}&2.551&1.0&1.44&4.57&$-1.55 \pm 0.19$&NA&NA&NA&4\\
{H1413+117B,C}&2.551&1.0&1.44&5.62&$-2.48 \pm 0.23$&NA&NA&NA&4\\
{H1413+117B,D}&2.551&1.0&1.44&3.97&$-2.38 \pm 0.23$&NA&NA&NA&4\\
{H1413+117C,D}&2.551&1.0&1.44&3.77&$0.10 \pm 0.19$&NA&NA&NA&4\\
{H1413+117A,B}&2.551&1.0&1.66&2.19&$0.60 \pm 0.19$&NA&NA&NA&4\\
{H1413+117A,C}&2.551&1.0&1.66&2.49&$-0.40\pm 0.19$&NA&NA&NA&4\\
{H1413+117A,D}&2.551&1.0&1.66&3.16&$-0.70 \pm 0.19$&NA&NA&NA&4\\
{H1413+117B,C}&2.551&1.0&1.66&3.88&$-1.00 \pm 0.19$&NA&NA&NA&4\\
{H1413+117B,D}&2.551&1.0&1.66&2.74&$-1.30 \pm 0.19$&NA&NA&NA&4\\
{H1413+117C,D}&2.551&1.0&1.66&2.60&$-0.30 \pm 0.19$&NA&NA&NA&4\\
{Q0957+561A,B}&1.414&0.36&1.391&0.37&$-0.38 \pm 0.13$&$-0.11 \pm 0.019$&NA&NA&5\\
{Q1104-1805A,B}&2.319&0.729&1.662&4.51&$-3.28 \pm 0.10$&$-2.30 \pm 0.08$&NA&NA&6\\
{UM673A,B}&2.743&0.493&1.626&2.72&$<-2.6$&NA&NA&NA&7\\
{SDSS J1442+4055}&2.590&0.284&1.946&0.69&$0.07 \pm 0.05$&$-0.13 \pm 0.042$&NA&NA&8\\
\enddata
\tablecomments{{ 1. Transverse separation in kpc between the sight lines at the absorber redshift. \\2. Linear  metallicity gradient in dex kpc$^{-1}$. 3. [Fe/H] inferred to be at the center of the \\lens galaxy (see text for more details). 4. References: (1) This work, (2) Zahedy et al. \\ (2017), (3) Lopez et al. (2005), (4) Monier et al. (2009), (5) Churchill et al. (2003), \\(6) Lopez et al. (1999), (7) Cooke et al. (2010), (8) Krogager et al. (2018).
5. Both \\ potential lens redshifts are listed for this GLQ. }
}
\label{tab_complit}
\end{deluxetable}
\end{landscape}
					
\end{document}